\documentclass{aa}
\usepackage{graphicx}

  \def\ngc{NGC~1692}
  \def\pks{PKS~0625--35}
  \def\3c8{3C~88}
  \def\3c4{3C~444}
  \def\oiii{[{\sc O\, iii}]}
  \def\oii{[{\sc O\, ii}]}
  
  \def\feka{Fe K$\alpha$}
  \def\chandra{{\it Chandra}} 
  \def\xmm{{\it XMM-Newton}} 
   
  \def\hst{{\it HST}} 
   
  \def\sax{{\it BeppoSAX}} 
   
  \def\rosat{{\it ROSAT}}

  \def\nh{cm$^{-2}$}
  \def\arcsec{$^{\prime\prime}$}
  \def\oi{\relax \ifmmode {\rm O\,{\sc i}}\else O\,{\sc i}\fi}
  \def\oii{\relax \ifmmode {\rm O\,{\sc ii}}\else O\,{\sc ii}\fi} 
  \def\oiii{\relax \ifmmode {\rm O\,{\sc iii}}\else O\,{\sc iii}\fi} 
  \def\ltsima{$\; \buildrel < \over \sim \;$}
  \def\simlt{\lower.5ex\hbox{\ltsima}} 
  \def\gtsima{$\; \buildrel > \over \sim \;$}
  \def\simgt{\lower.5ex\hbox{\gtsima}} 

\begin{document}

\title{The polyhedral nature of LINERs:\\
 An XMM-Newton view of LINERs in radio galaxies}

\author{M. Gliozzi\inst{1} 
\and L. Foschini\inst{2}  
\and  R.M. Sambruna\inst{3} 
\and F. Tavecchio\inst{4}} 
\offprints{mario@physics.gmu.edu} 
\institute{George Mason University, 4400 University Drive, Fairfax, VA 22030
\and INAF/IASF-Bologna, via Gobetti 101, 40129 Bologna, Italy
\and NASA Goddard Space Flight Center, Code 661, Greenbelt, MD 20771
\and INAF/Osservatorio Astronomico di Brera, via Bianchi 46, Merate and via
Brera 28, Milano, Italy }

\date{Received: ; accepted: }

\abstract 
{}
{We investigate the origin of X-rays and the  nature of accretion flow
in 4 low-ionization nuclear emission-line regions (LINERs) hosted by 
radio galaxies, namely \ngc, \pks, 3C~88, 3C~444, recently observed with \xmm.}
{We combine the results from the time-averaged spectral analysis
with model-independent information from X-ray temporal and spectral 
variability analyses, and with additional broadband information (specifically 
from the UV band, covered by
the Optical Monitor aboard \xmm, and from archival radio data).}
{The values of the Eddington ratios $L_{\rm bol}/L_{\rm Edd}$ of our 
sample span 2 orders of
magnitude ranging between $\sim 1\times 10^{-5}$ and $1\times 10^{-3}$.
The 4 AGN are adequately fitted by the same continuum model that
comprises at least one thermal component ($kT\sim 0.65-1.45$ keV)
and a partially absorbed power law, whose relative contribution and 
photon index vary substantially from source to source. \ngc\ and \pks\ have 
fairly steep power-law 
components ($\Gamma\sim 2.5-2.9$), perhaps indicative of synchrotron emission 
from the base of a jet. Conversely, the flat photon index derived for 3C~88 
($\Gamma\sim 1.1$) may be indicative of a heavily absorbed object. Finally,
the time-averaged spectral properties of \3c4 ($\Gamma \sim1.9$ and an 
apparent line-like excess around 6.7 keV) are more in line with Seyfert-like
objects. The temporal analysis reveals that
\pks\ and 3C~88 are significantly variable in the soft (0.2--1 keV) 
energy band. \pks\ also shows suggestive evidence of spectral variability on 
timescales of months, with a spectral softening associated with the source 
brightening. \ngc\ is only marginally variable in the soft band, whereas \3c4\
does not show significant variability on short timescales. The main findings
from the broadband analysis can be summarized as follows: 1) \3c4, \pks, and
\ngc\ have $\alpha_{OX}$ values consistent with the  $\alpha_{OX} - l_{UV}$ 
correlation found by Steffen et al. (2006) for Seyfert-like objects. 2) No
positive correlation is found between $L_{\rm X}$ and the inclination angle,
suggesting that the X-ray emission is not beamed. 3) The values of the
radio-loudness are inversely proportional to the Eddington ratio and locate
our objects in between the ``radio-loud'' and ``radio-quiet'' branches in the
$R- l_{UV}$ plane proposed by Maoz (2007).
}
{} 

\titlerunning{The polyhedral nature of LINERs}
\authorrunning{M.~Gliozzi et al.}
\maketitle

\section{Introduction}
The recent discovery of \hst\ and ground-based observations that most 
nearby galaxies harbor supermassive black holes (e.g., Magorrian et al. 1998; 
Gebhardt et al. 2003), provides convincing 
evidence that active galactic nuclei (AGN) and normal galaxies are 
fundamentally connected. It is now generally accepted that there is not a 
sharp distinction between AGN and normal galaxies, but rather a continuous 
distribution of activity levels between the two extremes. Therefore, of 
crucial importance is the role played by low-power AGN, because they may 
represent the link between powerful AGN and normal galaxies. 
\begin{table*}[htb]
\caption{ Source properties}
\begin{center}
\begin{tabular}{llcccccc}
\hline
\hline
\noalign{\smallskip}
Source &  Class  &   z  & $  N_{\rm H,Gal}$ & $ \log M_{\rm BH} $  &  $S_{\rm core,5GHz}$&  $S_{\rm tot,5GHz}$ &  $\Re$ \\    
\noalign{\smallskip}
       &         &      & $[{\rm 10^{20}~cm^{-2}}]$ &$[M_{\odot}]$ & [Jy]  & [Jy]     &  $[S_{\rm core}/S_{\rm ext}]$ \\     
\noalign{\smallskip}
\hline
\noalign{\smallskip}
\noalign{\smallskip}
  NGC~1692       & FRI &  0.035 &3.54 & 9.01 & 0.04 & 1.81 & 0.023 \\
\noalign{\smallskip}
\hline
\noalign{\smallskip}
 PKS~0625-35   & FRI &  0.055 &7.08 & 9.19 & 0.60 & 2.12 & 0.395 \\
\noalign{\smallskip}
\hline
\noalign{\smallskip}
3C~88        & FRI/FRII & 0.030  &8.81 & 8.70 & 0.16 & 2.14 & 0.087 \\
\noalign{\smallskip}
\hline
\noalign{\smallskip}
3C~444        & FRII& 0.153  &2.61 & 9.28 & $<$0.002& 1.99 & $<$0.0009\\
\noalign{\smallskip}
\hline
\hline
\end{tabular}
\end{center}
\footnotesize
{\bf Columns Table 1}: 1= Source name. 2= Fanaroff--Riley classification. 
3= Redshift.
4= Galactic column density. 5= Black hole mass estimate from Bettoni et al. (2003); the uncertainty is $\sim$0.4 dex. 6= Core radio flux from Morganti et al. (1993). 7= Total  radio flux from Morganti et al. (1993). 8= Radio core dominance, $S_{\rm core}/(S_{\rm tot}-S_{\rm core})$, from Morganti et al. (1993).
\label{tab1}
\end{table*}

The relevance of low-power AGN is bolstered by the finding that 
they make up a very substantial fraction of the
local galaxy population. Indeed, according to the Palomar survey 
(Ho et al. 1997), over 40\% of nearby 
galaxies contain low-power  AGN, and are optically classified as 
low-ionization nuclear emission-line regions (LINERs) or as ``transition
objects'', i.e. objects with optical spectra intermediate between those of
pure LINERs and H II regions (Heckman 1980). 

In the past, several studies investigated the nature of the ionizing mechanism
that produces LINER-like optical spectra (i.e., [\oii]/[\oiii]$>$1 and 
[\oi]/[\oiii]$>$0.33) and yielded different viable solutions. For example, 
Heckman (1980) and Dopita \& Sutherland (1995) proposed shock-heated gas as 
main ionizing mechanism. Alternatively, the ionization can be provided by 
very hot stars (such as Wolf-Rayet stars) clustered around the host galaxy 
nucleus (e.g., Terlevich \& Melnick 1985). Finally, it has been proposed that 
LINERs can be powered by low-power AGN (e.g., Filippenko \& Halpern 1984).

Currently, the dominant ionization mechanism responsible for these unusual 
line ratios is still a matter of debate. Nevertheless, it is generally accepted
that a sizable fraction of LINERs is associated with AGN.
This conclusion is supported by the discovery of broad emission lines in
many LINER spectra (e.g., Eracleous \& Halpern 2001), by X-ray imaging and 
spectral studies (e.g., Terashima \& Wilson 2003; Dudik et al. 2005), and by 
the fact that LINERs are often associated with compact radio sources with
high brightness temperatures (Nagar et al. 2005). 

However, the nature
of the AGN producing LINER-like spectra is still controversial. For example, based
on the study of non-simultaneous spectral energy distributions (SEDs) of 
seven LINERs, Ho
(1999) outlines the lack of a ``big blue bump'' (usually interpreted as
signature for standard efficient accretion disks), arguing that LINERs have 
intrinsically different SEDs compared to ``normal'' AGN and that they are
likely hosts of radiatively inefficient accretion flows (RIAFs). On the other
hand, Satyapal et al. (2004) combining mid-IR spectroscopy with X-ray imaging
conclude that LINERs are often heavily absorbed and that the intrinsic SEDs 
in LINERs are not necessarily different from those of standard AGN. This
conclusion finds support from the recent findings of Maoz (2007): Using 
high-quality UV data for a sample of 13 unabsorbed LINERs observed with \hst,
Maoz argues that the SEDs of low-luminosity AGN (LLAGN) are similar to 
those of Seyfert galaxies.

Previous X-ray studies of low-power AGN with \chandra\ demonstrate the 
importance 
of X-rays to make progress in this field (e.g.,
Di Matteo et al. 2001,2003; Pellegrini et al. 2005; Evans et al. 2006; 
Balmaverde et al. 2006). However, the nature of 
the central engine in low-power AGN is still a matter of debate, with some
authors favoring a jet-dominated scenario and others an accretion-dominated
scenario, which in turn might be either radiatively efficient or inefficient.
One of the reasons for this controversy is the poor discriminating
power of the low signal-to-noise (S/N) spectra, which do not allow 
one to choose between the competing scenarios.

In order to break this 
spectral degeneracy, higher S/N spectra combined with additional 
model-independent constraints are required. This is the approach that 
we (Gliozzi et al. 2003) adopted to investigate 
the origin of the X-rays in the nuclear region of the nearby 
FRI NGC~4261, which is classified as  Weak-Line Radio Galaxy (WLRG) 
and LINER (Lewis et al. 2003). By definition, WLRGs are radio galaxies with
very weak [\oiii] emission: EW([\oiii])$<$10\AA) (Tadhunter et al. 1998). 
However, no formal 
constraints are placed on the optical line ratios, therefore, an object
classified as WLRG is not necessarily a LINER and vice versa. Nevertheless,
a systematic study of the optical properties of a sample WLRGs has shown that 
the majority of the objects can be robustly classified as LINERs (Lewis et al.
2003). For the sake of simplicity, in the following we will use the term WLRGs
to indicate radio galaxies  with weak optical lines, which are optically 
classified as LINERs.

Exploiting the superior sensitivity of \xmm, in NGC~4261 we 
found evidence for significant temporal and spectral variability 
on short timescales, and the presence of an unresolved Fe K line at $\sim6.9$
keV of nuclear origin. We thus concluded that the inner jet
is unlikely to dominate the X-ray emission. 
Similar conclusions were reached by Rinn et al. (2005), using \chandra, \xmm, 
and \sax\ archival data of 9 WLRGs. 
Another relevant result from Rinn et al. (2005) is the detection of 
intrinsic absorption in 6 out 9 objects, with 4 sources having
$N_{\rm H}> 10^{22}$ \nh.
This is in contrast with the systematic 
lack of absorption found in a larger sample of non--LINER FRIs 
observed with \chandra\ or \xmm\
(Donato et al. 2004), and suggests that 
WLRGs may form a distinct class compared to other low--power radio
galaxies. 

To summarize, there is still no clear consensus on what is the dominant 
source of X-rays in low-power AGN/LINERs and what is the nature of 
accretion flow 
onto their black holes. Are the different conclusions just a consequence 
of the lower photon statistics, or are they reflecting 
intrinsic differences among objects of the same class? 
And if so, are these differences related to the nuclear power and/or 
to the beaming of the sources? To answer these questions and make progress 
in the study of accretion in AGN, it is important to expand the sample of 
low-power radio galaxies with high-quality X-ray data.

Here, we investigate the nature of  4 WLRGs (optically classified as 
LINERs)
\ngc, \pks, 3C~88, and 3C~444, whose properties are described in $\S$2
and summarized in Table 1.
Specifically,  using observations from \xmm\ (described in $\S$3), we try 
to shed light on 
the origin of X-rays and the nature of the accretion mechanism. To this end,
we combine several pieces of information obtained from the time-averaged
X-ray spectroscopy (discussed in $\S$4), from the X-ray variability (described
in $\S$5), and from broad-band analysis, which combines the simultaneous
X-ray/UV coverage with archival radio data (in $\S$6).
Hereafter, we adopt $H_0=71{\rm~km~s^{-1}~Mpc^{-1}}$, $\Omega_\Lambda=0.73$ and
$\Omega_{\rm M}=0.27$ (Bennet et al. 2003).

\section{Sample Properties}
The 4 targets, whose properties are reported in Table 1, were selected 
from a sample of 20 WLRGs recently studied in the optical range by Lewis
et al. (2003). This sample was in turn drawn from the sample of
WLRGs presented by Tadhunter et al. (1998). The main characteristics of
the 4 radio galaxies can be summarized as follows:
\begin{itemize}
\item 
They all contain a confirmed LINER on the basis of measurements of optical
line ratios (Lewis et al. 2003). 

\item Two (\ngc\ and \pks) are FRI and two (3C~88 and \3c4) FRII galaxies, 
although the classification of 3C~88 is only morphological. 
According to the optical criteria
proposed by Jackson \& Rawlings (1997), the two FRII galaxies are classified
as low-excitation radio galaxies (LEGs). 

\item They have been previously detected at X-rays in the \rosat\ all sky 
survey; \pks\ was also observed with \sax\ (Trussoni et al. 1999). The X-ray 
luminosities span two orders of magnitude. 

\item They are all nearby galaxies ($z=0.03-0.15 $) well 
studied at longer wavelengths, especially in the radio band with ATCA and VLA
observations.

\item The masses of their supermassive black holes have been estimated
in an homogeneous way by Bettoni et al (2003). 
\end{itemize}

In summary, although the radio core fluxes and the morphological classification
are different, the 4 sources share several fundamental common properties: 
1) Optically, they are all formally classified as LINERs. 2) In the radio 
band, they have similar 
extended fluxes, which, unlike the core fluxes (possibly affected by beaming), 
are generally considered reliable indicators of the source intrinsic power.
3) They have similar black hole masses, which are indistinguishable from
each other when their uncertainty of the order of 0.4 dex is accounted for.
4) Recent radio, optical, and 
X-ray studies of the core properties of radio galaxies have suggested that 
FRII/LEGs are indeed unified with FRI galaxies (e.g., Chiaberge et al. 2002; 
Hardcastle et al. 2006).

\section{Observations and Data Reduction}
We observed our 4 targets with \xmm\ between August, 2005 and September,
2006. The nominal durations range between 15 and 35 ks.
All of the EPIC cameras (Str\"uder et al. 2001; Turner et al. 2001) were 
operated in full-frame mode with  thin 
or medium filters, depending on the presence of bright nearby sources in the
field of view. As a precaution, for \pks\ and \3c4\ the MOS cameras were 
operated in small window mode to prevent photon pile-up.
The recorded events were screened to remove known
hot pixels and other data flagged as bad; only data with {\tt FLAG=0}
were used.  
\begin{figure*}
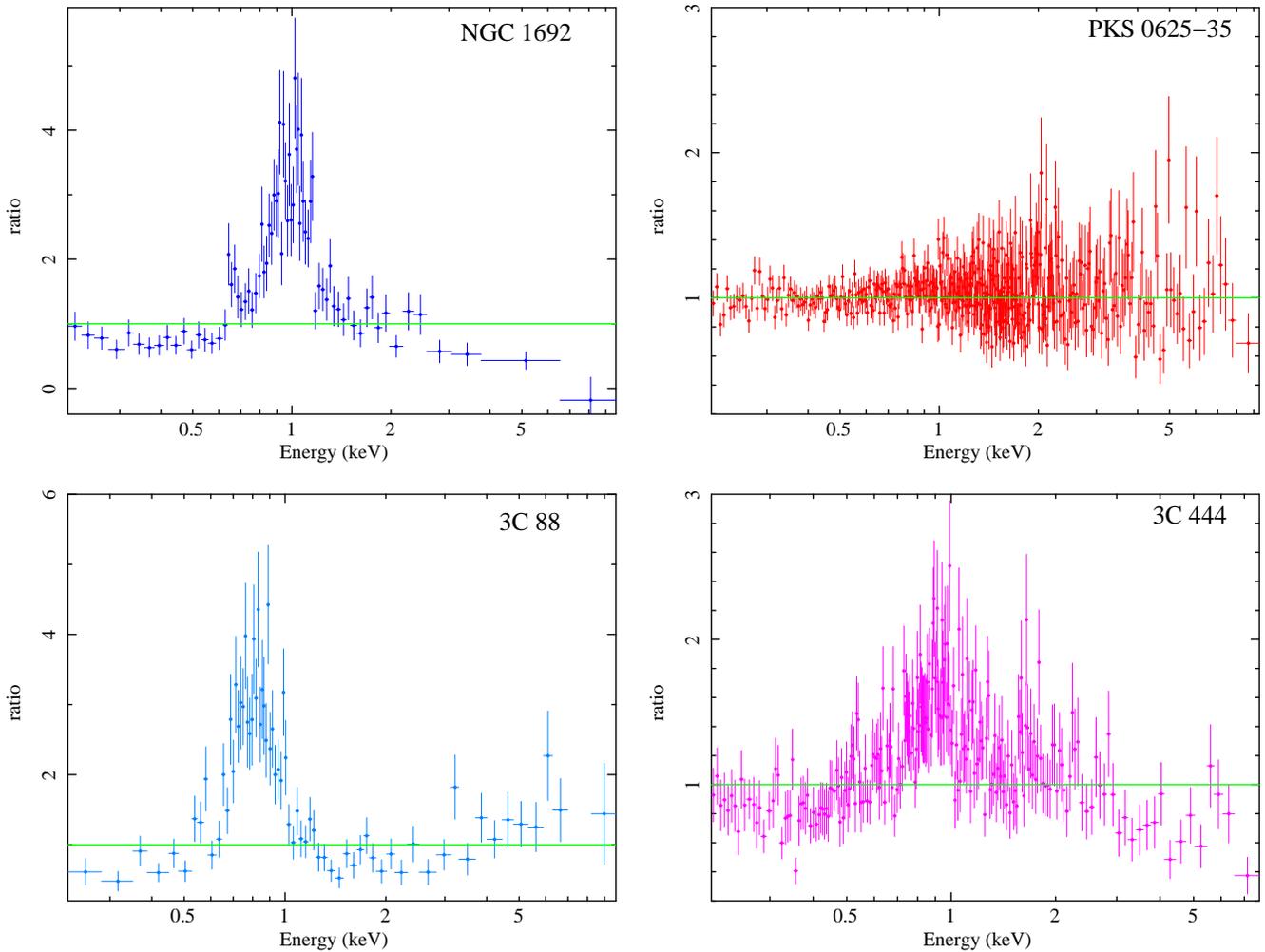

\includegraphics[bb=60 10 595 720,clip=,angle=-90,width=9.cm]{f1a.eps}\includegraphics[bb=60 10 595 720,clip=,angle=-90,width=9.cm]{f1b.eps}
\includegraphics[bb=60 10 595 720,clip=,angle=-90,width=9.cm]{f1c.eps}\includegraphics[bb=60 10 595 720,clip=,angle=-90,width=9.cm]{f1d.eps}
\caption{Data-to-model ratio obtained by fitting the 
EPIC pn spectra in the 0.2--10 keV energy range with a simple power-law
model absorbed by Galactic $N_{\rm H}$. Large residuals at soft energies
are present in \ngc, 3C~88, and \3c4, which also shows an excess around
6 keV. \pks\ is apparently  adequately described by a simple power law. 
However, a more detailed analysis suggests more complex spectral models
also for this source (see text for details). 
}
\label{figure:fig1}
\end{figure*}

The data were processed using the latest CCD gain
values. For the temporal analysis, events corresponding to pattern 
0--12 (singles, doubles, triples, and quadruples) in the MOS cameras and 
0--4 (singles and doubles only, since the pn pixels are larger) in the pn 
camera were accepted. For the spectral analysis, only single events were
considered in order to exploit their better calibration down to 0.2 keV
(Kirsch 2006). Arf and rmf files were created with 
the \xmm\ Science Analysis Software (\verb+SAS+) 7.0.
Investigation of the full--field light curves revealed the
presence of one period of background flaring for \ngc, and several flares
for the EPIC pn observation of 3C~88.
These events were excluded, reducing the effective
total exposures time to the values reported in Table 2. 
 The extraction radii used for source
spectra and light curves range between  20\arcsec\ (for 3C~88)
and 40\arcsec\ (for 3C~444), depending on the extension of the source and 
its proximity to the CCD edges.  For reference, with the chosen
cosmological parameters, 10\arcsec\ correspond to 6.9 kpc for \ngc,
10.5 kpc for \pks, 5.9 kpc for 3C~88, and 26.3 kpc for \3c4.
Background spectra and light curves were extracted from
source-free circular regions on the same chip as the source, with 
extraction radii $\sim$2 times larger than those used for the source. 
 There are no
signs of pile-up in the pn or MOS cameras according to the {\tt SAS}
task {\tt epatplot}.  
The RGS data  have signal-to-noise ratio ($S/N$) that is too low for a 
meaningful analysis.

The observation log with
dates of the observations, EPIC net exposures, and the average count rates
are reported in Table 2. 
\begin{table}[ht]
\caption{ Observation Log}
\begin{center}
\scriptsize
\begin{tabular}{lccc}
\hline
\hline
\noalign{\smallskip}
Source          &  Date  &  MOS exposure  &   MOS rate       \\    
\noalign{\smallskip}
                    & [dd/mm/yyyy] &   [ks]             & [${\rm s^{-1}}$]  \\
\noalign{\smallskip}
\hline
\noalign{\smallskip}
\noalign{\smallskip}
  NGC~1692     & 15/08/2005   & 13.8             & $(6.7\pm0.2)\times 10^{-2}$         \\
\noalign{\smallskip}
\hline
\noalign{\smallskip}
   PKS~0625-35 & 03/05/2005   &  1.5               & $(7.4\pm0.2)\times 10^{-1}$       \\ 
               & 27/08/2005   & 13.1               & $(8.1\pm0.1)\times 10^{-1}$             \\
\noalign{\smallskip}
\hline
\noalign{\smallskip}
 3C~88       & 02/09/2006   & 22.4               & $(3.4\pm0.2)\times 10^{-2}$             \\   
\noalign{\smallskip}
\hline
\noalign{\smallskip}
 3C~444       & 02/11/2005   & 18.1               & $(1.2\pm0.1)\times 10^{-1}$            \\   
\noalign{\smallskip}
\hline
\hline
\end{tabular}
\end{center}
\footnotesize
{\bf Columns Table 2}: 1= Source name. 2= Date of observation. 3= EPIC MOS2 net exposure.
4= EPIC MOS2 net count rate in the 0.2--10 keV band (the MOS1 count rate is consistent
with the MOS2 value within the quoted errors, whereas the pn count rate is $\sim$2.5 times the
MOS count rate). 
\label{tab2}
\end{table}

The spectral analysis  was performed using the {\tt XSPEC v.12.3.0}
software package (Arnaud 1996; Dorman \& Arnaud 2001). 
The EPIC data have been re-binned in order to contain at least 20
counts per channel, depending on the brightness of the source.
The errors on spectral parameters are at 90\% confidence level for one 
interesting parameter ($\Delta \chi^2=2.71$).

The data from the OM (Mason et al. 2001) were processed with the latest 
calibration files using the {\tt SAS}
task {\tt omichain}, which provides  count rates in the U (3440\AA),
UVW1 (2910\AA), and UVM2 (2310\AA) bands. The count rates were then
converted into magnitudes using the formula 
${\rm mag=-2.5\log(count~rate)+zero-point}$ (where the zero-points are 18.259,
17.204, and 15.772, for the U, UVW1, and UVM2 bands, respectively), and
corrected for systematics. The values were 
corrected for extinction (see $\S$6 for
more details) and converted to fluxes 
by using standard formulae (e.g., Zombeck 1990).

\section{X-ray Spectral Analysis}
The X-ray spectra represent one of the most used and effective tools
to investigate the accretion process in AGN. In the following, 
we perform a time-averaged spectral analysis of the EPIC data
in the 0.2--10 keV energy range. For the analysis of the continuum, the 
spectra from the three EPIC cameras (pn, MOS1, and MOS2) have been combined
to improve the overall S/N.  The EPIC pn alone was used for the \feka\ analysis
in virtue of its larger throughput and effective area above 6 keV. 
For the sake of clarity, in the figures describing the results
from the spectral analysis we show only the EPIC pn data.

\begin{figure*}
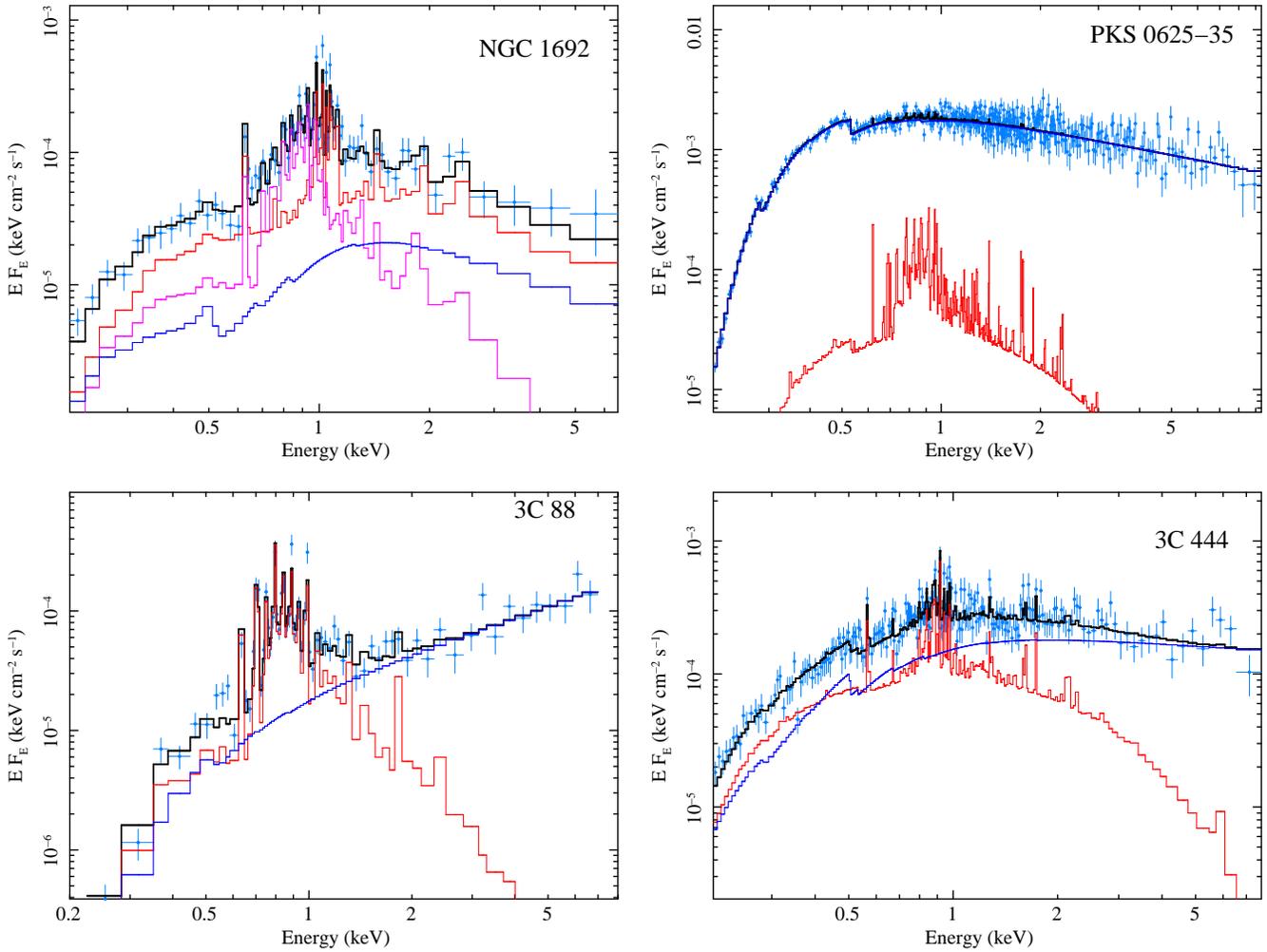

\includegraphics[bb=60 10 595 720,clip=,angle=-90,width=9.cm]{f2a.eps}\includegraphics[bb=60 10 595 720,clip=,angle=-90,width=9.cm]{f2b.eps}
\includegraphics[bb=60 10 595 720,clip=,angle=-90,width=9.cm]{f2c.eps}\includegraphics[bb=60 10 595 720,clip=,angle=-90,width=9.cm]{f2d.eps}
\caption{Deconvolved EPIC pn spectra in the 0.2--10 keV energy range,
fitted with a partially absorbed power-law plus a thermal component.
\ngc\ requires an additional thermal component, whereas \3c4\ shows a
line-like excess around $\sim$ 6-7 keV. All models are absorbed by
Galactic $N_{\rm H}$. 
}
\label{figure:fig2}
\end{figure*}

\begin{figure*}
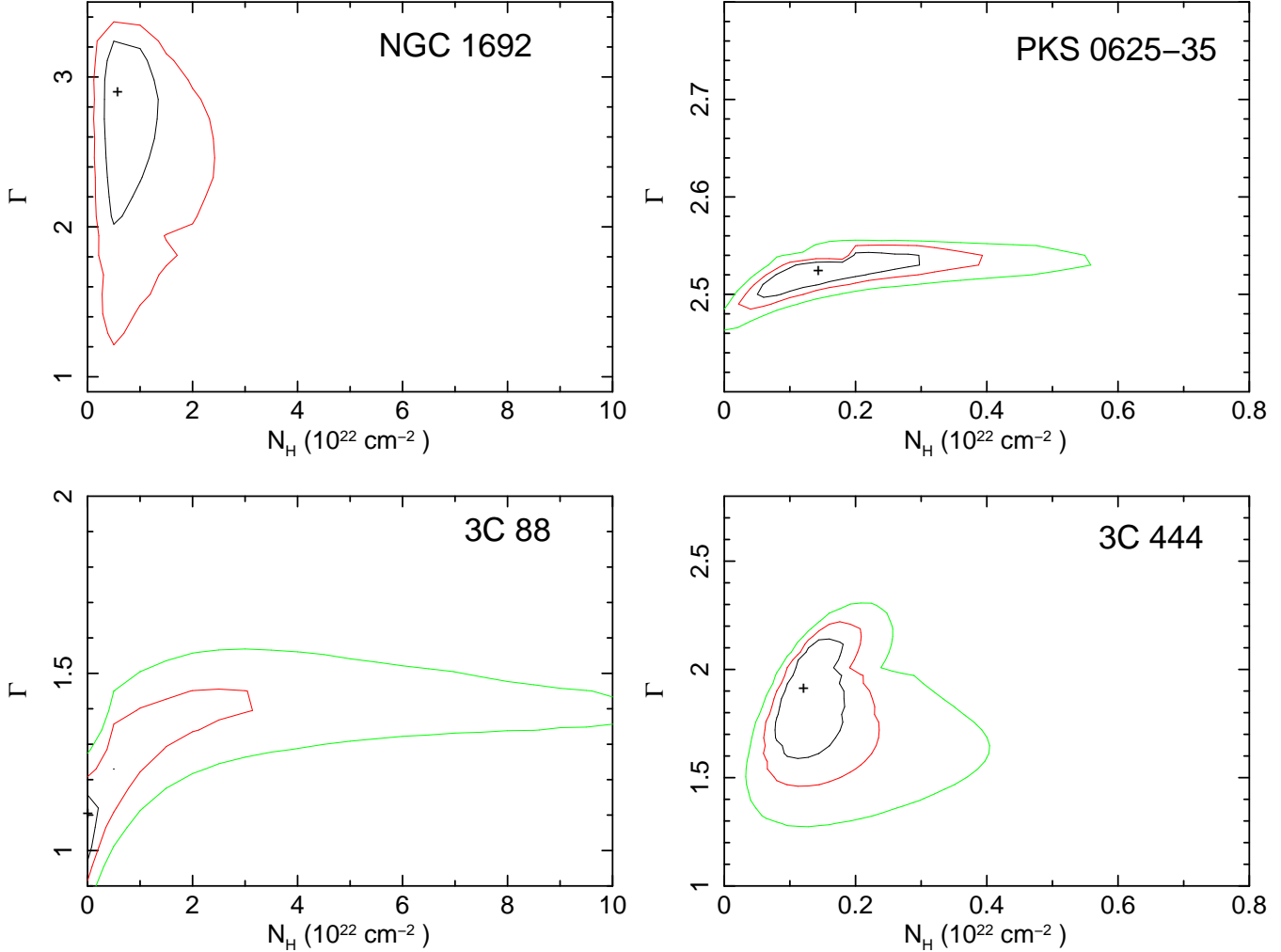

\includegraphics[bb=120 95 545 645,clip=,angle=-90,width=9.cm]{f3a.eps}\includegraphics[bb=120 95 545 645,clip=,angle=-90,width=9.cm]{f3b.eps}
\includegraphics[bb=120 95 545 645,clip=,angle=-90,width=9.cm]{f3c.eps}\includegraphics[bb=120 95 545 645,clip=,angle=-90,width=9.cm]{f3d.eps}
\caption{Confidence contours (68\%, 90\%, and 99\%) in the $\Gamma$ --
$N_{\rm H}$ plane; for \ngc, only 68\% and  90\% levels are shown.
The spectral model used comprises a thermal component and a partially
absorbed power law.}
\label{figure:fig3}
\end{figure*}

\subsection{Continuum}
The background-subtracted EPIC spectra in the 0.2--10 keV energy range were
fitted with models of increasing complexity until an adequate fit was achieved.
The significance of the different spectral components was obtained from an
$F-$test. As an operative rule, if the addition of a spectral component 
improves the fit by $\Delta\chi^2$ such that
the associated $P_F\leq 0.05$, the new component is considered significant.

In general, simple power-law models, modified by Galactic 
interstellar absorption, do not provide a good representation of the \xmm\ 
EPIC spectra.
 Figure~\ref{figure:fig1} shows the data-to-model ratios for our 4 targets.
Large residuals at soft energies are present in \ngc, 3C~88, and \3c4, 
suggesting the presence of additional (thermal) components. The residuals of 
\3c4\ also suggest the possible presence of a line-like feature around 6--7 
keV. Only \pks\ is apparently  well described by a simple power law.  
However, a more detailed analysis indicates that more 
complex spectral models provide better fits also for this source (see below).

\begin{table*}[ht]
\caption{X-ray continuum spectral properties}
\begin{center}
\begin{tabular}{lccccccccc}
\hline
\hline
\noalign{\smallskip}
Source &  \multicolumn{2}{c}{ Thermal Component} && \multicolumn{3}{c}{Partial Absorption} && Power Law & $\chi^2$(dof) \\    
\noalign{\smallskip}
       &      $kT$    & $P_F$                    && $  N_{\rm H}$  & $Cvr$ & $P_F$ && $\Gamma$  &\\
       &      [keV]   & [\%]                     &&$[{\rm 10^{21}~cm^{-2}}]$ & [\%]&[\%]&&    &\\     
\noalign{\smallskip}
\hline
\noalign{\smallskip}
\noalign{\smallskip}
  NGC~1692  &  $0.79\pm0.08$ & $3\times10^{-9}$      && $6_{-4}^{+12}$ & 95  & $2\times10^{-2}$&& $2.9_{-1.1}^{+0.4}$ & 131.6(143)\\
\noalign{\smallskip}
\hline
\noalign{\smallskip}
 PKS~0625-35&  $0.64\pm0.19$ & $3\times 10^{-3}$  && $1.5_{-1.1}^{+1.8}$ & 12 & $1\times10^{-2}$ && $2.52_{-0.03}^{+0.02}$   & 791.3(762) \\
\noalign{\smallskip}
\hline
\noalign{\smallskip}
3C~88        & $0.64\pm0.02$ & $1\times 10^{-15}$      && $0.5_{-0.5}^{+6.2}$ & 50  & $5\times10^{-1}$ &&$1.11_{-0.14}^{+0.18}$&  164.6(133)\\
\noalign{\smallskip}
\hline
\noalign{\smallskip}
3C~444       & $1.45\pm0.15$ &$7\times 10^{-7}$       && $1.2_{-0.5}^{+0.7}$ & 95 & $6\times 10^{-5}$       && $1.92_{-0.33}^{+0.27}$ & 394.4(377)\\
\noalign{\smallskip}
\hline
\hline
\end{tabular}
\end{center}
\footnotesize
{\bf Columns Table 3}: 1= Source name. 2= Temperature ($kT$) of the  thermal component. 
3=  Probability associated with the $F$ statistic (the smaller the value the higher the significance
of the spectral component).
4= Intrinsic column density of a partial covering model. 5= Covering factor. 6= Probability associated with 
the $F$ statistic for the partially absorbed model. 7= Photon index. 
8= $\chi^2$ and number of degrees of freedom. {\bf Notes:} \ngc\ requires a second thermal component
with $kT=1.44\pm0.15$ ($P_F=5\times 10^{-9}$). The quoted uncertainties
correspond to 90\% confidence levels.
\label{tab3}
\end{table*}

The baseline model able to fit the spectra of the 4 targets requires at
least 2 components: 1) a thermal
component, described by \verb+apec+ in \verb+XSPEC+ (Smith et al. 2001),
which  parameterizes the emission from a collisionally-ionized 
plasma and is usually associated with a diffuse circum-nuclear component
that is frequently detected in radio galaxies by \chandra\ thanks to its 
sub-arcsecond spatial resolution (e.g., Donato et al.
2004), and 2) a partially
absorbed power law that emerges at harder energies and parameterizes the
emission from the unresolved nuclear component. The metal abundances of the
thermal model were initially left free to vary between 0.2 and 1.2 solar values, 
and then fixed at the best fit value. Also the covering factor in the partial
absorption model (\verb+zpcfabs+ in \verb+XSPEC+) was kept fixed at the best 
fit value.

We have also tried alternative models such as  broken power-law 
(hereafter bkn) or
double power-law models, with one absorbed power law with $\Gamma$ fixed
at 1.8 and the other power law unabsorbed. However, none of the alternative
models yielded  better (or statistically equivalent) fits compared to results 
from the baseline model described 
above. In the case of the double power-law model, the spectral parameters 
are in general poorly constrained and often the contribution of one of the 
two power laws becomes negligible. 
The only exception is \pks, for which a 
broken power law ($\Gamma_1=2.35\pm0.06$, $E_{\rm br}=0.8\pm0.3$ keV, 
$\Gamma_2=2.54\pm0.03$)  provides an equally good statistical 
fit using less free parameters than
the baseline model ($\chi^2=790.1$ for 765 dof and $\chi^2=791.3$ for 762 dof,
for  bkn and the baseline model, respectively). However, since the 
fluxes and luminosities are fully consistent with those obtained using the
partially absorbed model, for consistency and to allow a direct comparison 
with the other sources, in Table 3 we report the  results of the spectral 
fitting of the continua (and the significance of the individual components)
of the baseline model described above. 

The photon indices obtained for our
sample vary substantially from source to source, ranging from $\Gamma\sim 1.1$
for 3C~88  to $\Gamma\sim2.5-2.9$ for \pks\ and \ngc\ respectively, with 
\3c4\ being the only one to show a typical Seyfert-like $\Gamma\sim 1.9$.
All the sources statistically
require at least one thermal component with $kT$ in the range 0.65--1.45 keV.

\begin{table*} 
\caption{X-ray Fluxes and Luminosities}
\begin{center}
\begin{tabular}{lccccccccc}
\hline
\hline
\noalign{\smallskip}
Source & $F_{\rm 0.2-2~keV}$ & \multicolumn{2}{c}{$L_{\rm 0.3-2~keV}$} && $F_{\rm2-10~keV}$ & \multicolumn{2}{c}{$L_{\rm2-10~keV}$} && $L_{\rm bol}/L_{\rm Edd}$  \\
\noalign{\smallskip}
       &   $[{\rm erg~cm^{-2}~s^{-1}}]$ &   $[{\rm 10^{42}~erg~s^{-1}}]$   & \%  $L_{\rm PL}$  &&  $[{\rm erg~cm^{-2}~s^{-1}}]$ &  $[{\rm 10^{42}~ erg~s^{-1}}]$    &  \%  $L_{\rm PL}$ &&\\
\noalign{\smallskip}
\hline
\noalign{\smallskip}
\noalign{\smallskip}
  NGC~1692  & $2.7\times10^{-13}$&  0.9 & 55\% && $0.9\times10^{-13}$&   0.2 & 39\% && $1\times10^{-5}$\\
\noalign{\smallskip}
\hline
\noalign{\smallskip}
 PKS~0625-35 & $4.7\times10^{-12}$ &  78.2 & 98\% && $2.6\times10^{-12}$ &  18.7 & 100\% && $2\times10^{-3}$    \\
\noalign{\smallskip}
\hline
\noalign{\smallskip}
3C~88     & $1.4\times10^{-13}$& 0.4 & 34\% &&  $2.8\times10^{-13}$& 0.5 & 99\% && $2\times10^{-4}$    \\
\noalign{\smallskip}
\hline
\noalign{\smallskip}
3C~444   & $6.8\times10^{-13}$ &63.0 & 50\% && $5.1\times10^{-13}$ &  33.1 & 78\% && $3\times10^{-3}$       \\
\noalign{\smallskip}
\hline
\hline
\end{tabular}
\end{center}
\footnotesize
{\bf Columns Table 4}: 1= Source name. 2= Soft (0.2--2 keV) X-ray absorbed flux. 
3= Soft X-ray intrinsic luminosity. 4= Percentage of the soft X-ray luminosity 
associated with the power-law component.
5= Hard (2--10 keV) X-ray absorbed flux.
6= Hard  X-ray intrinsic luminosity. 7= Percentage of the hard X-ray luminosity 
associated with the power-law component. 8= Eddington ratio $L_{\rm bol}/L_{\rm Edd}$,
where $L_{\rm bol}$ was
obtained by multiplying the 2--10 keV X-ray luminosity, associated 
with the power law, by a factor 20 (see text for details). 
\label{tab4}
\end{table*}       

Although, the baseline model
is the same for the 4 sources, the contribution of the different components
changes significantly for the different objects, as it is clearly shown in
Figure~\ref{figure:fig2}, where the deconvolved $E~F_{\rm E}$ spectra are 
plotted. In addition, \ngc\ 
statistically requires the presence of a second thermal component.

One of the goals of the X-ray spectral analysis is to investigate the presence
of intrinsic absorption. Table 3 already indicates that there is evidence 
for moderate absorption ($N_{\rm H}\sim10^{21}$\nh) in \ngc, \pks, and \3c4,
whereas in 3C~88 a local absorber is not statistically required. However, 
the latter result might be hampered by the limited photon statistics, since 
the unusually flat photon index may indicate that the X-ray spectrum of 3C~88 
is affected by substantial absorption. Indeed, if $\Gamma$ is fixed at a more
reasonable value of 1.4, the resulting fit ($\chi^2=168.6$ for 134 dof) is 
still comparable to the one obtained with $\Gamma$ free, however, the resulting
intrinsic absorption, $N_{\rm H}=1.8_{-1.4}^{+3.7}\times10^{22}$\nh,  is now 
substantial. 

In order to test whether there
is degeneracy between  $\Gamma$ and $N_{\rm H}$, and to better
visualize the relative uncertainties on these parameters, in 
Figure~\ref{figure:fig3} we have 
plotted the confidence contours of $\Gamma$ and  $N_{\rm H}$. 
For the sources with the highest count rate, \pks\ and \3c4,  
$N_{\rm H}$ is well constrained, whereas the plots of \ngc\ and 3C~88 show 
large uncertainties,
which do not allow one to rule out a heavily absorbed scenario.

In Table 4 we report the values for the soft (0.2--2 keV) and hard (2--10 keV)
fluxes and luminosities as well as the relative contributions of the 
power-law component in these energy bands.
In general, the hard energy band is dominated by the power-law component with
contributions larger than $\sim$80\%, the only exception being \ngc, where the 
power law contributes only  $\sim$40\% of the total $L_{2-10~\rm keV}$. 

Using the
estimates of the black hole masses reported in Table 1 and the
hard X-ray luminosities associated with the power law (hereafter $L_{\rm X}$),
we have computed the luminosity ratio $L_{\rm X}/L_{\rm Edd}$ (where $L_{\rm Edd}$ 
is the Eddington luminosity). Detailed modeling 
of broad-band SEDs of weak AGN and Galactic black holes in the low/hard state
showed that the origin of X-rays (accretion-dominated versus jet-dominated)
may indeed depend upon the value of $L_{\rm X}/L_{\rm Edd}$, with 
values below a threshold of $10^{-6}$ generally associated with a jet-dominated
scenario (see Wu et al. 2007 and references therein for more details). 
In this framework, 
the relatively high values of 
$L_{\rm X}/L_{\rm Edd}$ derived for \pks\ ($1\times10^{-4}$), 3C~88
($1\times10^{-5}$), and \3c4\ ($1\times10^{-4}$) apparently favor an
accretion-dominated origin, whereas the relatively low value found for \ngc\
($8\times10^{-7}$) is more in line with a jet origin. However, one must
bear in mind that the above criterion is strongly model-dependent and hence
these conclusions should be taken with caution. 

In general, the accretion efficiency in black hole systems is assessed by
computing the Eddington ratio $L_{\rm bol}/L_{\rm Edd}$, which requires the
use of a bolometric correction factor. Often in AGN studies this factor 
has been 
computed by assuming a mean energy distribution, which cannot be 
accurate for all AGN (Elvis et al. 1994). A further problem with this 
approach is that the contribution from the IR bump is included in the
computation of the bolometric luminosity. However, the IR
emission in AGN is generally dominated by reprocessed emission and hence 
its inclusion
would lead to an overestimate of the accretion luminosity.
More recently, Marconi et al. (2004)
proposed an alternative approach, based on Montecarlo simulations of a 
template of AGN SEDs, which yields different bolometric correction factors
depending on the AGN X-ray luminosity. Finally, Vasudevan \& Fabian (2007),
using actual SEDs from 54 AGN, showed that the  dependence on the X-ray 
luminosity is not so straightforward, although there seems 
to be a bimodal dependence on the Eddington ratio: for
$L_{\rm bol}/L_{\rm Edd}<$0.1 the 
correction factor to $L_{\rm 2-10~keV}$ is 15-25, while above this threshold
the correction factor is 40-70. 

Following  Vasudevan \& Fabian (2007) and bearing in mind several caveats
(for example, the above results are derived from non-simultaneous SEDs of 
radio-quiet AGN), we computed the bolometric luminosity
for our targets by multiplying $L_{\rm 2-10~keV}$ by a correcting factor 
of 20 (i.e., we assume that the 2-10 keV luminosity represents 5\% of the 
bolometric luminosity). The results, reported in the last column of 
Table 4, indicate that the Eddington ratio for the 4 targets 
spans more than two orders of magnitude ranging between 
$\sim 1\times 10^{-5}$ and $3\times 10^{-3}$.
While the lowest values (obtained for \ngc\ and 3C~88)
seem to favor a radiatively inefficient accretion flow hypothesis (e.g.,
Quataert et al. 1999), 
a radiatively efficient scenario cannot be ruled out a priori for \pks\ and
\3c4, if we assume that the bulk of the X-ray emission is produced in the
accretion flow. These conclusions hold even if the uncertainties for the
black hole masses are accounted for, given that they cause changes of
$L_{\rm bol}$ by a factor of a few and not by orders of magnitude.

\begin{figure}
\includegraphics[bb=118 100 525 645,clip=,angle=-90,width=9.cm]{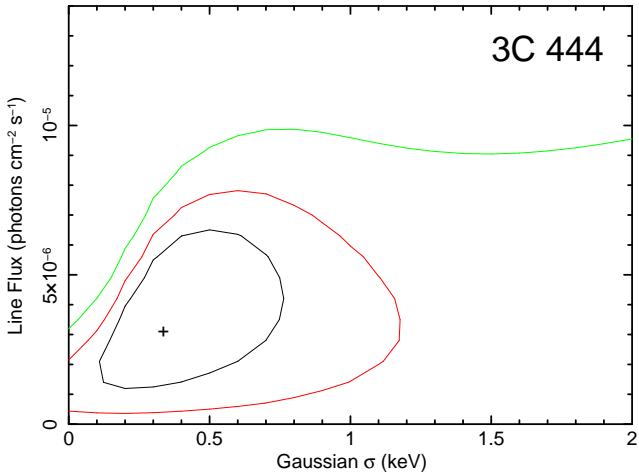}
\caption{ Confidence contours in the line intensity--line width
plane for \3c4, using a single Gaussian model. The confidence levels 
correspond to 68\%, 90\%, and 99\%. 
 }
\label{figure:fig4}
\end{figure}

\subsection{Iron K$\alpha$ line} 
\begin{table*} 
\caption{Fe K$\alpha$ Gaussian Model Parameters (Rest Frame)}
\begin{center}
\begin{tabular}{lclccc} 
\hline        
\hline
\noalign{\smallskip} 
Source  & $E_{\rm Fe}$  &$\sigma_{\rm Fe}$ &  $I_{\rm Fe}$                       & EW    &    $\Delta\chi^2$ (dof)  \\ 
        &  (keV)        &   (keV)          &  ($10^{-6}{\rm~ph~cm^{-2}~s^{-1}}$) & (keV)  &      \\
\hline 
\noalign{\smallskip}
PKS~0625-35       &  $6.4$  &   0.01   & $1.5_{-1.5}^{+2.4}$        & 0.079 ($<$0.182)  & -1.3 (1)\\      
\noalign{\smallskip}
\hline
\noalign{\smallskip}
3C~88         &  $6.4$  &   0.01   & $1.5_{-1.5}^{+5.5}$        & 0.435 ($<$0.643)  & -1.5 (1)\\      
\noalign{\smallskip}
\hline
\noalign{\smallskip}
3C~444        &  $6.7_{-0.4}^{+0.4}$ & $0.35_{-0.35}^{+0.47}$ & $4.4_{-2.7}^{+2.8}$    & $0.550_{-0.380}^{+0.690}$  &  
-5.7 (3)\\      
\noalign{\smallskip}
\hline
\hline
\end{tabular}
\end{center}
\footnotesize
{\bf Columns Table 5}: 1= Source name. 2= Line energy in the source's rest frame. 
3= Line width. 4= Line flux. 5= Equivalent width. 6= Decrease in the $\chi^2$ (and additional
number of degrees of freedom) obtained by adding a Gaussian line to the best fit continuum.
\label{tab5}
\end{table*}     

After finding adequate parameterizations for the spectral continua, we have 
investigated whether a \feka\ line could improve the overall fits. An 
inspection of the residuals does not show evidence for any prominent line-like features in the 6--7 keV range for \pks and 3C~88 (\ngc\ has no spectral 
coverage of the 
interesting energy range for this analysis). Nevertheless, we have added
a narrow ($\sigma=0.01$ keV) Gaussian line with energy fixed at 6.4 keV (in the 
source's rest frame)
to the best-fit continuum models, in order to estimate the 90\% upper limit
on the line contribution, as measured by the equivalent width. The results
are summarized in Table 5.

Unlike the other sources, the spectral residuals of \3c4\ around 6--7 keV
apparently reveal a relatively broad hump centered around $\sim$6.7 keV (in the 
source's rest frame). Adding a Gaussian line
to the best-fitting continuum model improves the fit of the 
EPIC pn spectrum only marginally: $\Delta\chi^2\sim -6$ for 3 additional 
parameters (however, if the line energy is kept fixed at 6.7 keV, the line 
significance increases at more than 90\%).

When fitted with a single Gaussian, the line appears to be 
centered around 6.7 keV,  possibly broad
($\sigma\sim0.35$ keV) and fairly strong ($EW\sim550$ eV). Similar results
are obtained when a diskline model 
(with inclination angle fixed at 60\degr) is 
used instead of a Gaussian line. 
We have also tried to fit the line-like feature with two Gaussian models,
keeping one line energy fixed at 6.4 keV (in the source's rest frame).
The overall fit does not improve significantly ($\Delta\chi^2=
-1.5$ for 2 additional dof), and 
the results, $E_1=6.4$ keV, $\sigma_1\ll 10^{-3}$ keV,
$EW<440$ eV, $E_2=7.0_{-0.7}^{+0.5}$ keV, $\sigma_1=0.2_{-0.2}^{+0.7}$ keV,
$EW=475_{-99}^{+744}$ eV, indicate that the line parameters are poorly
constrained.

The contour plots of the line flux and physical width (see 
Fig.~\ref{figure:fig4}) obtained using a single Gaussian model,
apparently confirm that the line is detected at more than  90\%
level and that it is resolved at more than 68\% confidence level. 
However, a broader
hard X-ray energy range covering the region where the reflection component
peaks, and 
longer observations
with at least $10^4$ counts in the 2--10 keV energy band are necessary to
firmly establish the presence of a broad line and constrain the line parameters
(Guainazzi et al. 2006).

\section{X-ray Variability}
Although the time-averaged spectral analysis offers important insights into
the physical processes at work in AGN, it cannot be considered an exhaustive 
diagnostic. This is because the X-ray spectral analysis is based on the 
comparison of the observed spectrum with a model spectrum (the spectral fit), 
and hence the results are by definition model-dependent. 
In addition, the same spectra can often be adequately fitted by different 
combinations of spectral parameters for a
specific model, or by different physical models. This
spectral degeneracy severely affects spectra with low S/N (either due to
the intrinsic faintness of the sources or to a limited sensitivity of the
instrument), but also higher quality spectra are not immune from this problem.
In order to progress in our understanding of the AGN phenomenon, the results
from the X-ray time-averaged spectral analysis need to be validated and
complemented by other constraints (possibly obtained in a model-independent 
way) such as temporal and spectral variability studies.
\xmm\ with its high throughput and highly elliptical orbit is one of the
best instruments to investigate the AGN variability on short and medium
timescales.

The importance of combining information from the spectral and temporal
analyses is underscored by the power of X-ray spectral 
variability studies in discriminating between jet-dominated and
accretion-dominated emission. 

\begin{figure*}
\begin{center}
\includegraphics[bb=95 20 440 465,clip=,angle=0,width=9cm]{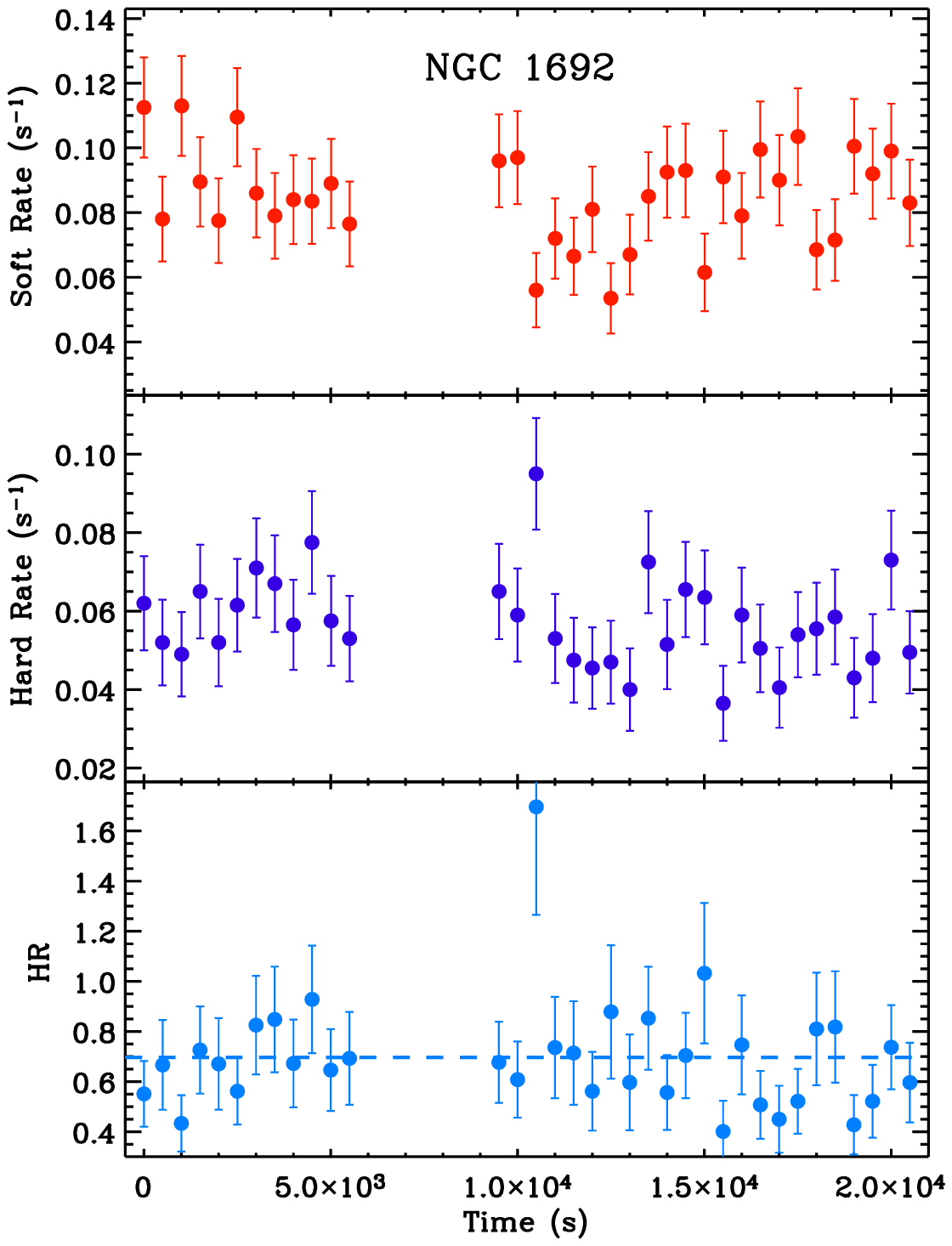}\includegraphics[bb=95 20 440 465,clip=,angle=0,width=9cm]{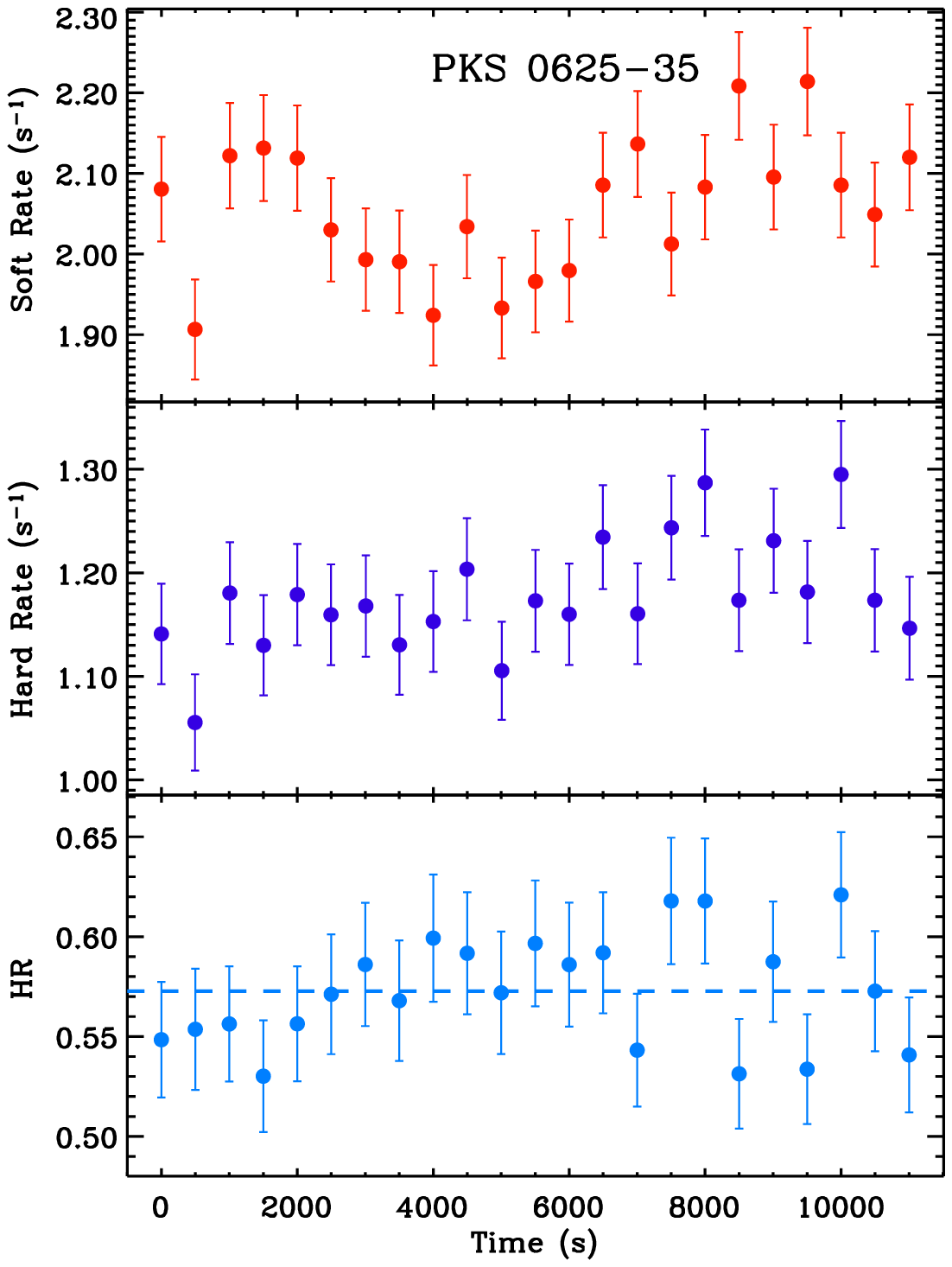}
\includegraphics[bb=95 20 440 465,clip=,angle=0,width=9cm]{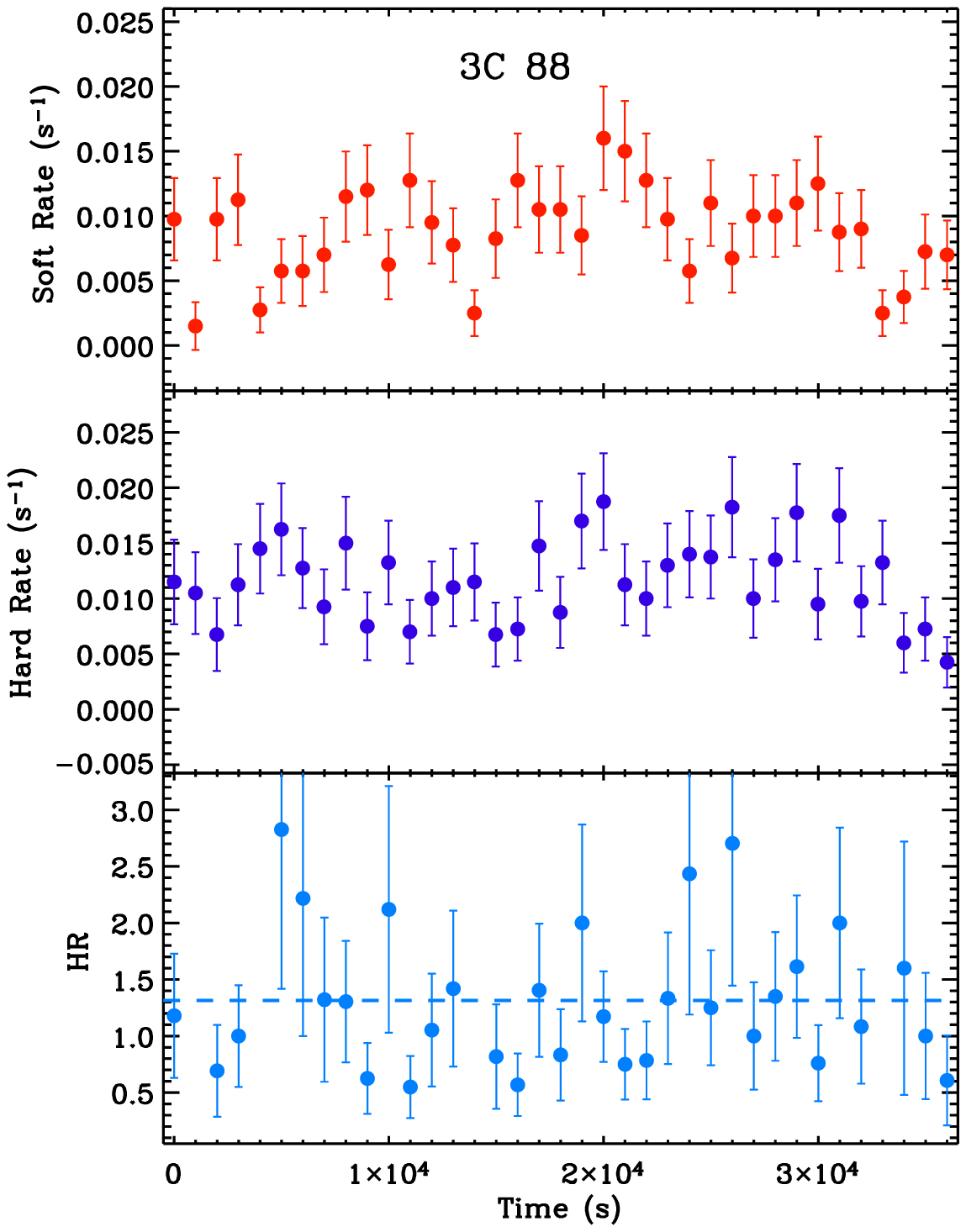}\includegraphics[bb=95 20 440 465,clip=,angle=0,width=9cm]{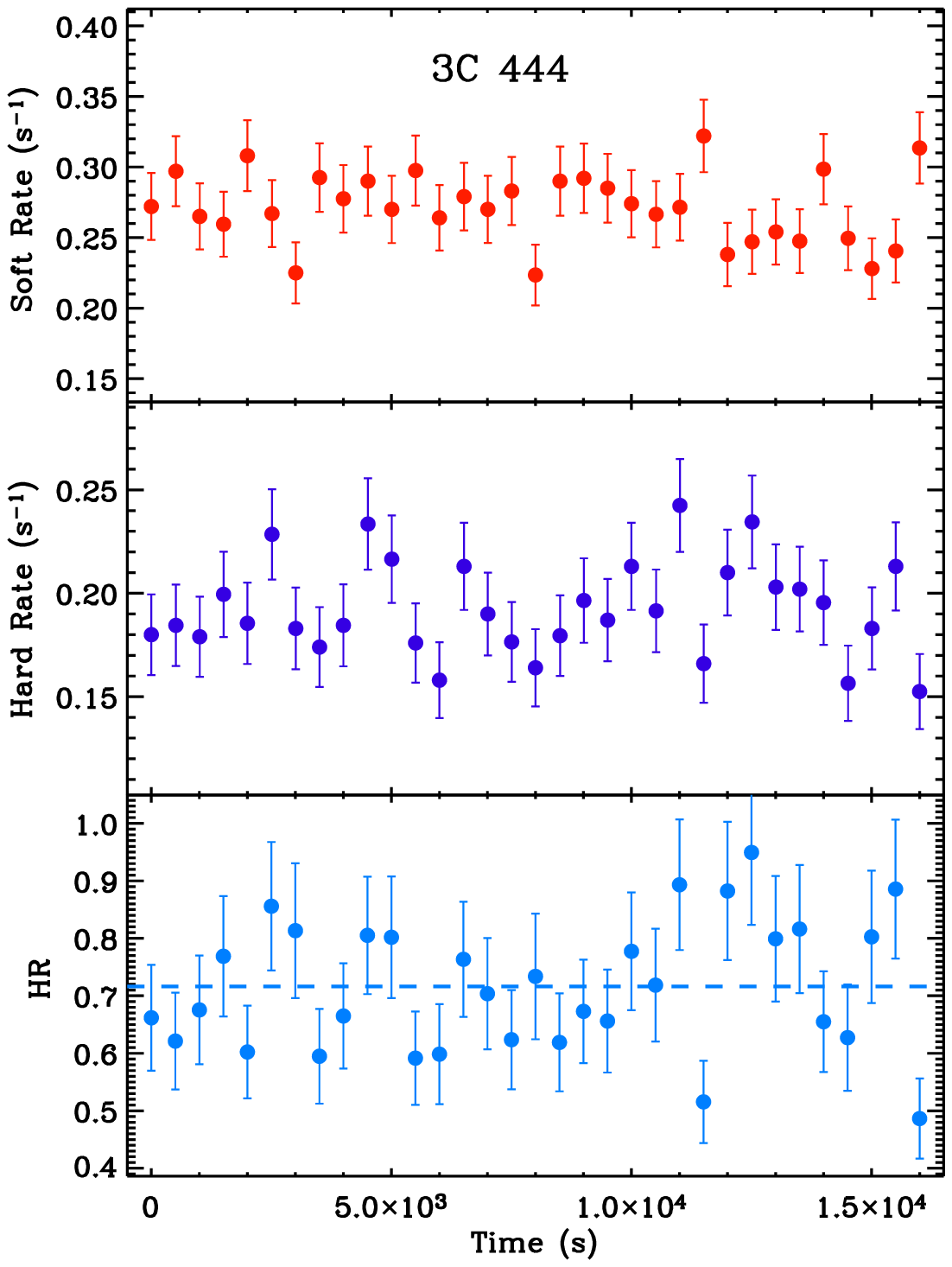}
\end{center}
\caption{\xmm EPIC light curves in the 0.2--1 keV (top panels),
1--10 keV (middle panels) energy bands, and HR=(1--10 keV/0.2--1 keV) 
plotted versus the time (bottom panels). For \ngc, \pks, and \3c4\ we have
used EPIC pn data and time bins of 500 s; for 3C~88, EPIC MOS2 data and 
time bins of 1000 s have been used.}
\label{figure:fig5}
\end{figure*}

For example, an X-ray spectrum described
by a power law with $\Gamma$ ranging between 1.5 and 2.5 can be equally
well interpreted in the framework of jet models (where inverse Compton
and synchrotron are the dominant emission processes), or in the framework 
of coronal
models, by varying the geometry configuration of static models (e.g.,
Haardt et al. 1994) or by assuming a mildly relativistic outflowing corona
(e.g., Malzac et al. 2001), or in the RIAF scenario by varying the relative
contributions of Bremsstrahlung and inverse Compton (e.g., Quataert et al. 
1999). 

Similarly, the presence of temporal X-ray variability alone is unable to rule 
out either scenarios. Indeed, continuous variability on all the sampled 
timescales is a common characteristic not only of blazars, but also of 
several classes of
radio quiet AGN (most notably, Seyfert and Narrow Line Seyfert 1 galaxies).

On the other hand, if both spectral and temporal information are combined,
a markedly different behavior is revealed by radio-quiet and jet-dominated
sources. Radio-quiet AGN typically show a softening of the spectrum that 
accompanies a flux increase (e.g., Papadakis et al. 2002; Markowitz et al. 
2003, and references therein), and a more pronounced variability in the 
soft  band than in the hard energy range (Markowitz \& Edelson 2001).
Conversely, blazars usually show an opposite spectral trend with 
a bluer when brighter behavior and with the fractional
variability more pronounced at higher energies (e.g., Fossati et al. 2000;
Gliozzi et al. 2006; Rebillot et al. 2006).

It must be noticed that intermediate BL Lac objects can sometimes mimic a
Seyfert-like spectral variability trend. This occurs when the bandpass
of the X-ray detector collects both soft X-rays associated with the
highly variable synchrotron component and hard X-rays associated with
the slowly variable inverse Compton component (e.g., Tagliaferri et
al. 2000; Foschini et al. 2007). However, this unusual (for jet-dominated
sources) spectral variability
trend is accompanied by an X-ray spectrum parameterized by a concave broken 
power-law model and hence easily discernible.

\subsection{Flux Variability} 
We studied the short-term variability using EPIC pn data with time-bins
of 500 s for \ngc, \pks, and \3c4. For 3C~88, since the EPIC pn data  
are severely
affected by background flaring, we have used EPIC MOS2 data with time-bins of
1000 s (given the lower count rate of the MOS data compared to the pn data).
The total 0.2--10 keV energy band was divided into a soft (0.2--1 keV) and a
hard (1--10 keV) bands, such that the mean count rates in the two sub-bands
are comparable and not noise-dominated.

Figure~\ref{figure:fig5} shows the EPIC time series of the 0.2-1 keV
(top panels), the 1-10 keV energy band (middle panels), and of the hardness 
ratio $HR=$1-10 keV/0.2-1 keV (bottom panels).
A visual inspection of Fig.~\ref{figure:fig5} suggests that low-amplitude
flux changes are a common phenomenon in the 4 radio galaxies. 
\begin{table*} 
\caption{X-ray Variability}
\begin{center}
\begin{tabular}{lccccccccccccc}
\hline
\hline
\noalign{\smallskip}
Source &&  \multicolumn{2}{c}{Total} && \multicolumn{2}{c}{Soft} &&  \multicolumn{2}{c}{Hard}  &  HR  \\
\noalign{\smallskip}
  &&$P_{\chi^2}$ &$F_{\rm var}$ && $P_{\chi^2}$ &$F_{\rm var}$ &&  $P_{\chi^2}$ &$F_{\rm var}$ & $P_{\chi^2}$  \\
\noalign{\smallskip}
\hline
\noalign{\smallskip}
\noalign{\smallskip}
  NGC~1692  && 0.30 & $0.03\pm0.07$ && 0.10 & $0.07\pm0.05$ && 0.45 & $0.05\pm0.10$ & 0.70\\
\noalign{\smallskip}
\hline
\noalign{\smallskip}
 PKS~0625-35 && 0.002 & $0.025\pm0.006$ && 0.02 & $0.027\pm0.009$ && 0.21 & $0.020\pm0.015$ & 0.59 \\
\noalign{\smallskip}
\hline
\noalign{\smallskip}
3C~88     && 0.10 & $0.10\pm0.07$ && $<0.001$ & $0.20\pm0.09$ && 0.10 & $0.09\pm0.13$ & 0.07    \\
\noalign{\smallskip}
\hline
\noalign{\smallskip}
3C~444    && 0.39 & $0.03\pm0.03$ && 0.23 & $0.06\pm0.03$ && 0.12 & $\dots$ & 0.02     \\
\noalign{\smallskip}
\hline
\hline
\end{tabular}
\end{center}
\footnotesize
{\bf Columns Table 6}: 1= Source name. 2= $\chi^2$ probability of constancy in the total 0.2--10 keV band. 
3= Fractional variability in the total 0.2--10 keV band.
4= $\chi^2$ probability of constancy in the soft 0.2--1 keV band. 
5= Fractional variability in the total 0.2--1 keV band. 
6= $\chi^2$ probability of constancy in the hard 1--10 keV band. 
7= Fractional variability in the total 1--10 keV band. 
8= $\chi^2$ probability of constancy in the HR=hard/soft light curve.
{\bf Note:} For NGC~1692, PKS~0625-35 and3C~444 the variability results have been obtained using EPIC pn light curves with
time bins of 500 s. For 3C~88, we have used EPIC MOS2 light curves with
time bins of 1000 s, because the EPIC pn data are severely affected by background flaring.
\label{tab6}
\end{table*} 

In order to
quantify the variability in the different energy bands, we have applied
$\chi^2$ tests and measured the fractional variability 
$F_{\rm var}={(\sigma^2-\Delta^2)^{1/2}/\langle r\rangle}$,
where $\sigma^2$ is the variance, $\langle r\rangle$ the
unweighted mean count rate, and $\Delta^2$ the mean square value of
the uncertainties associated with each individual count rate. 
Considering the brevity of the observations and the limited
count rates, as an operative rule, we deem a time series
significantly variable if $P_{\chi^2}<0.05$ and marginally variable if
$P_{\chi^2}=0.1$ (where $P_{\chi^2}$ is the probability  of constancy
associated with a $\chi^2$ test).

The results of the variability analysis are reported in Table 6 and
can be summarized as follows:\\
1) In the 0.2--10 keV energy band,
only \pks, which is the object with the highest count rate and the
least contribution from the thermal component, is significantly variable. 
3C~88 is marginally variable, whereas no variability is detected in
\ngc\ and \3c4.\\
2) In the soft band, both \pks\ and 3C~88 are significantly
variable, whereas \ngc\ is marginally variable, although the latter result
should be considered with caution given the low count rate and the
presence of a gap due to background flaring.\\ 
3) In the hard band, none of
the sources show significant variability, with the exception of 3C~88, which
is only marginally variable. The lack of variability in the hard band may be
partly related to the lower count rate. On the other hand, \pks\ (which has 
an average
hard count rate $>1.2{~\rm s^{-1}}$) suggests that this difference might be
intrinsic. Furthermore, because of the dilution from the thermal component,
the variability measured in the soft band should be considered as a lower 
limit, supporting the hypothesis that the sources are more variable at
softer energies.\\ 
4) There is some evidence for spectral variability
in 3C~88 and \3c4, suggesting a different temporal behavior in the two energy 
bands.
 
\subsection{Spectral Variability} 
\begin{figure}
\begin{center}
\includegraphics[bb=30 30 355 300,clip=,angle=0,width=9cm]{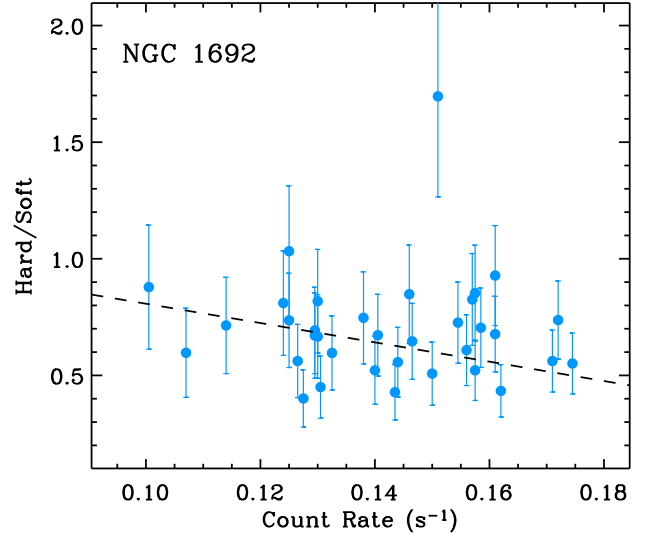}
\end{center}
\caption{Hardness ratio (1--10 keV/0.2--1 keV) plotted versus the count 
rate for \ngc. The gray (blue in color) filled circles correspond to time-bins
of 500 s. The dashed line represent the best-fit model obtained from a 
least-squares method. }
\label{figure:fig6}
\end{figure}

A useful method for investigating the nature of spectral variability
is based on the hardness ratio plotted versus the count rate. As an
example, Figure~\ref{figure:fig6}
shows the hard/soft X-ray color plotted versus the count rate
for \ngc, which is the source with the longest EPIC pn coverage.
The gray (blue in color) filled circles correspond to
time-bins of 500 s.  

\begin{table*} 
\caption{Optical and UV magnitudes}
\begin{center}
\begin{tabular}{llccc}
\hline
\hline
\noalign{\smallskip}
Source & filter ($\lambda$)& observed  & de-reddened    & de-reddened   \\
      &                    & magnitude & magnitude (SMC) & magnitude (X-ray)\\
\noalign{\smallskip}
\hline
\noalign{\smallskip}
\noalign{\smallskip}
NGC~1692  &U (3440\AA) & 16.4 & 16.2 & 13.1\\
\noalign{\smallskip}
         &UVW1 (2910\AA) & 19.4 & 19.1 & 16.0\\
\noalign{\smallskip}
         &UVM2 (2310\AA) &$>$22.0 &$>$21.6 &$>$18.5\\
\noalign{\smallskip}
\hline
\noalign{\smallskip}
 PKS~0625-35 & U (3440\AA) & 15.9 & 15.6 & 15.0\\
\noalign{\smallskip}
         &UVW1 (2910\AA) & 15.7 & 15.3 & 14.7\\
\noalign{\smallskip}
         &UVM2 (2310\AA) & 15.8 & 15.2 & 14.6\\
\noalign{\smallskip}
\hline
\noalign{\smallskip}
3C~88     &U (3440\AA) & 18.6 & 18.0 & 18.1\\
\noalign{\smallskip}
         &UVW1 (2910\AA) & 18.6 & 17.8 & 18.0\\
\noalign{\smallskip}
         &UVM2 (2310\AA) & 18.6 & 17.4 & 17.6\\
\noalign{\smallskip}
\hline
\noalign{\smallskip}
3C~444    &U (3440\AA) & 19.4 & 19.3 & 18.7\\
\noalign{\smallskip}
         &UVW1 (2910\AA) & 19.2 & 19.0 & 18.5\\
\noalign{\smallskip}
         &UVM2 (2310\AA) & 19.6 & 19.4 & 18.8\\
\noalign{\smallskip}
\hline
\hline
\end{tabular}
\end{center}
\footnotesize
{\bf Columns Table 7}: 1= Source name. 2=  OM filter and central wavelength .
3= Observed magnitude..
4= Magnitude de-reddened using a SMC-type extinction law (see text for details).
5= Magnitude de-reddened using a reddening  $A(V)/N_{\rm H}=5.3\times 10^{-22}$ with $N_{\rm H}$ derived from
the X-ray spectral fitting.
Note: For 3C~88 only the U filter yielded a value for the magnitude. The same magnitude was used for
the other 2 filters to allow a direct comparison with the other targets.  
\label{tab7}
\end{table*}

A visual inspection of Figure~\ref{figure:fig6}
suggests the presence of a weak negative trend with the source softening
when the count rate increases.  The dashed line represents the
best-fit model to the binned data point, obtained from a least-squares
method: $HR=(1.2\pm0.7) - (4.1\pm4.7)\; r$ (where $r$ is the count
rate). As indicated by the large errors,
the brevity of the observation and the low amplitude of
the flux variations hamper this kind of analysis, and no firm conclusions
can be derived on the presence of a trend in the $HR-ct$ plot. 
Nevertheless, compared to a constant
model (i.e., a model with slope=0 and intercept=$\langle HR \rangle$),
the best-fit linear model improves the data fit by $\Delta\chi^2\sim4$
for one additional degree of freedom.  
\begin{figure*}
\begin{center}
\includegraphics[bb=35 205 580 660,clip=,angle=0,width=9cm]{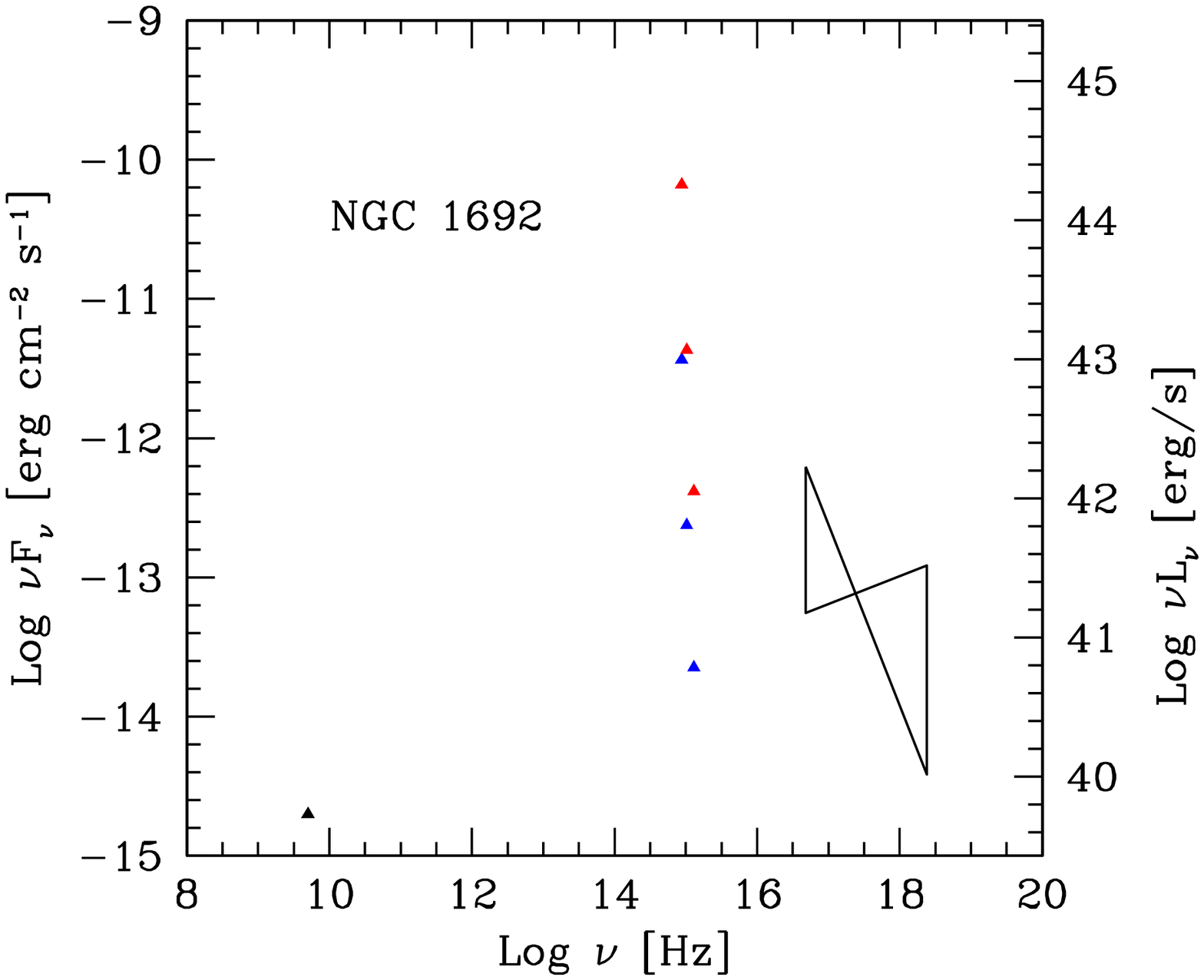}\includegraphics[bb=35 205 580 660,clip=,angle=0,width=9cm]{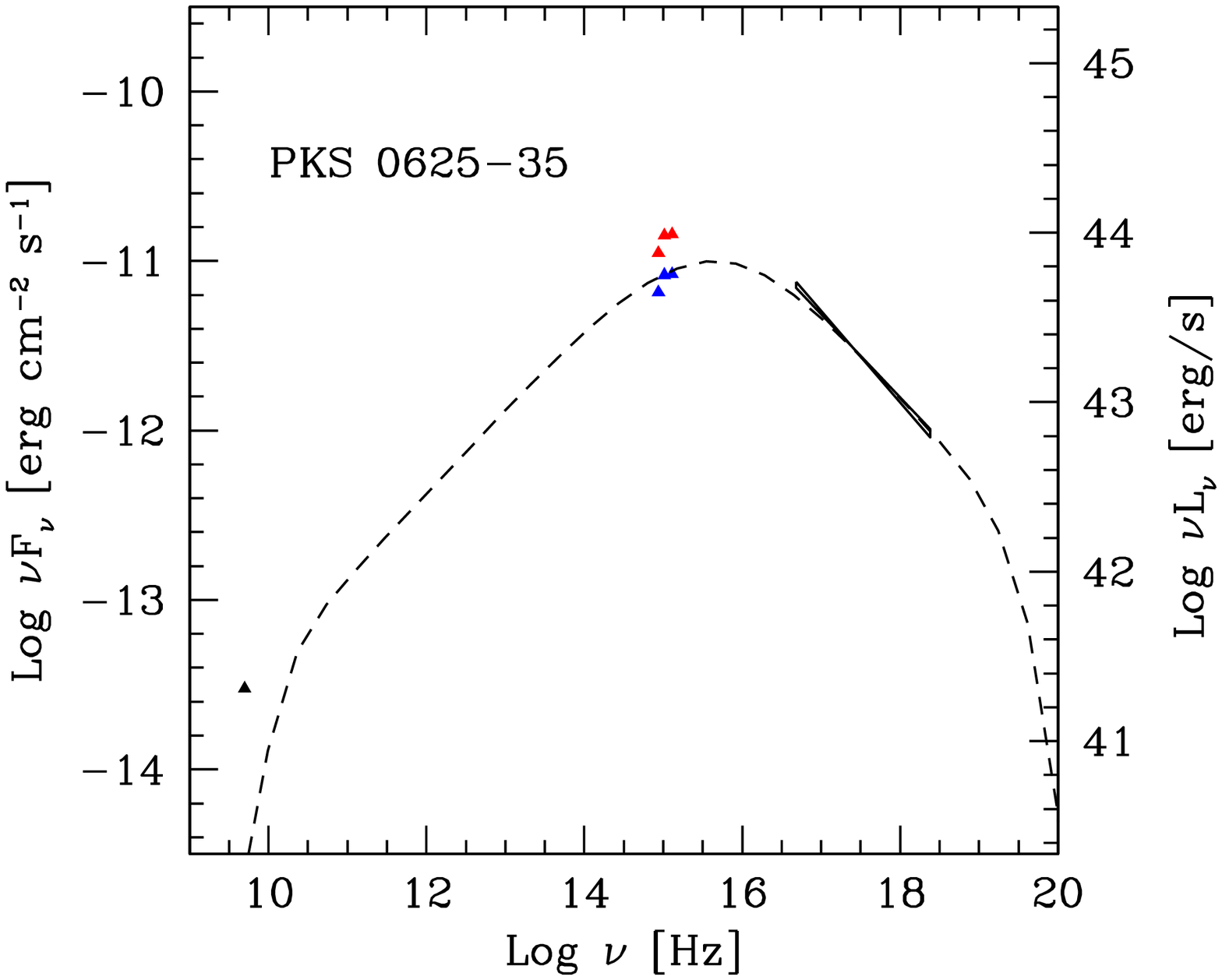}
\includegraphics[bb=35 205 580 660,clip=,angle=0,width=9cm]{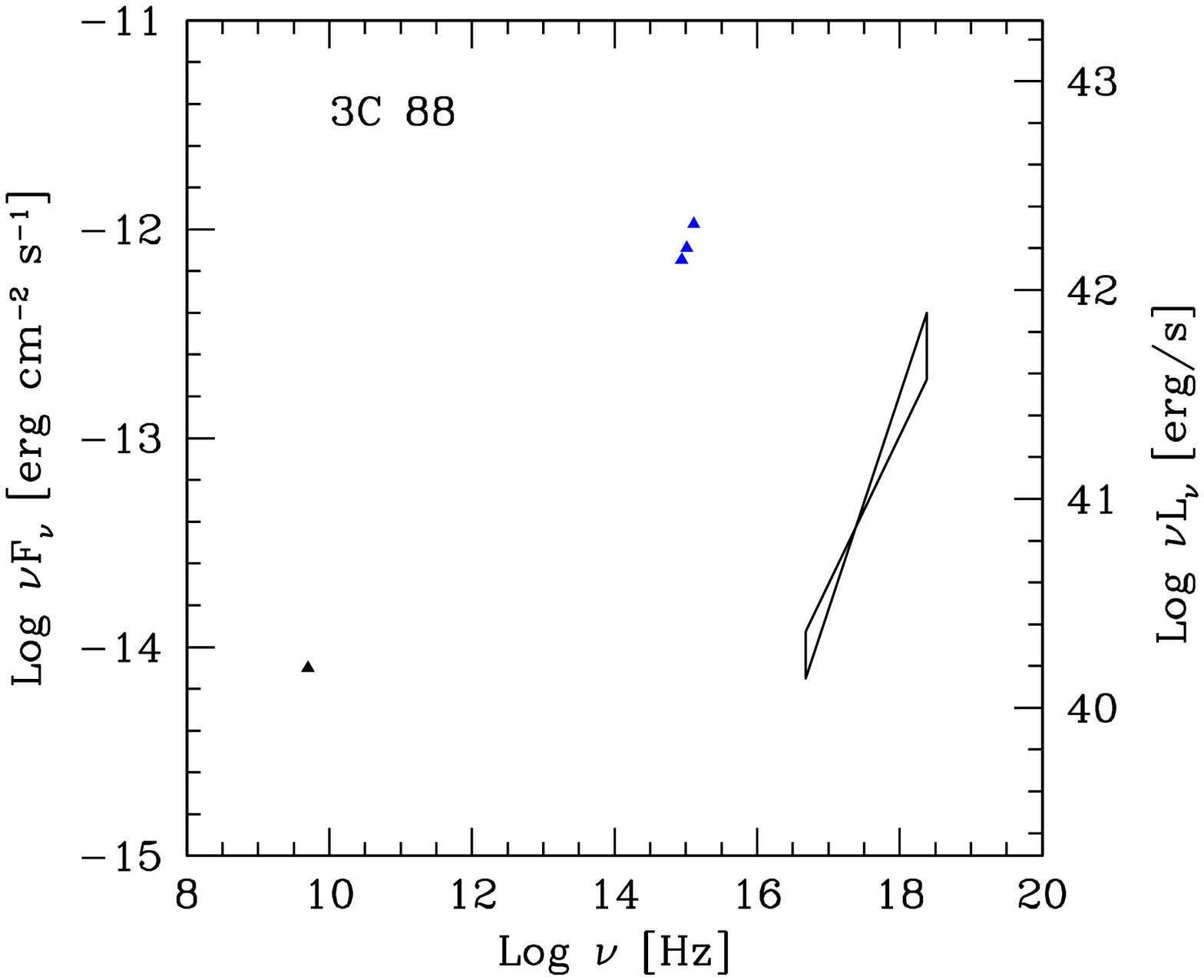}\includegraphics[bb=35 205 580 660,clip=,angle=0,width=9cm]{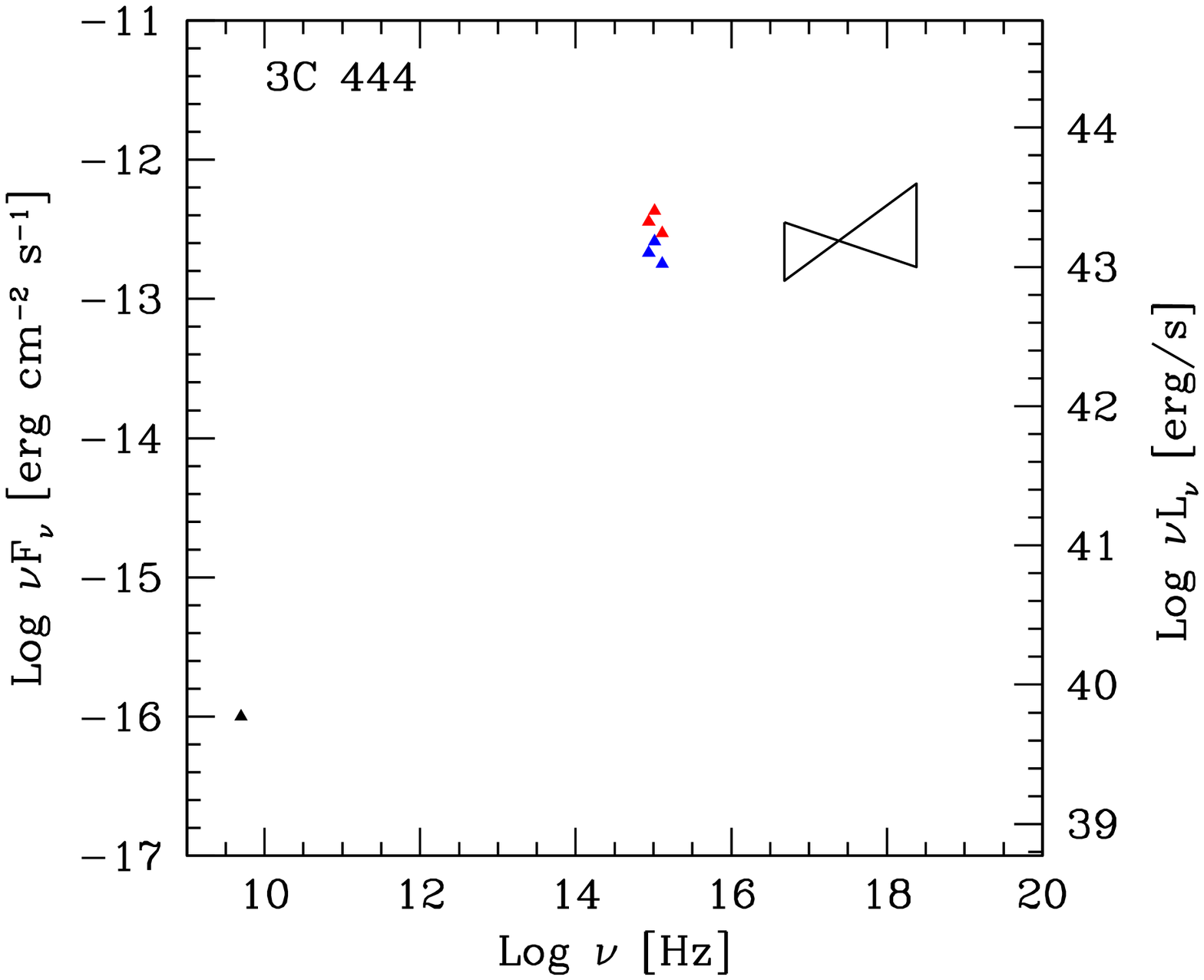}
\end{center}
\caption{Spectral energy distributions of the 4 radio galaxies/LINERs.
The optical/UV data are from the OM aboard \xmm\ and thus simultaneous
to the X-ray data. The radio data are taken from the literature.
The two sets of the optical/UV data (red and blue
in colors) for \ngc, \pks, and \3c4
correspond to two different extinction corrections (see text for more 
details). The dashed line superimposed to the \pks\ data represents the 
best fit from the jet model described in the text.}
\label{figure:fig7}
\end{figure*}

Since AGN are generally characterized by red-noise variability (i.e., the 
variability is more pronounced on longer timescales), in order
to successfully perform this kind of analysis, 
each source would require 
two observations  performed by the same instrument 
months/years a part, in order to increase the chances of observing substantial
flux and spectral changes. 

For technical reasons the exposure of \pks\ was split into two observations
(of duration of $\sim1.5$ ks and $\sim13$ ks, respectively) taken at beginning
of May, 2005 and toward the end of August, 2005. Although the shortness of the
first observation does not allow to properly constrain the spectral parameters
(due to their large uncertainties, they are consistent with those obtained
during the second observation), it yields enough counts to investigate the
spectral variability using the hardness ratio. However, since during the first
observation, the EPIC pn camera did not provide valid data, the comparison
between the hardness ratio has been performed using MOS data only.
For May 2005, when the average
EPIC MOS count rate was $r_{\rm t1}=0.74\pm0.02 {\rm~s^{-1}}$, the hardness 
ratio was
$HR_{\rm t1}=1.10\pm0.04$, whereas 4 months later the increase of the  average
count rate by $\sim$10\%, $r_{\rm t2}=0.81\pm0.01 {\rm~s^{-1}}$, was 
accompanied by an the hardness ratio $HR_{\rm t2}=0.97\pm0.01$,  yielding
an apparently significant decrease of $HR_{\rm t1}-HR_{\rm t2}=0.13\pm0.04$.
In other words,
the two \xmm\ observations suggest that \pks\ follows the typical 
Seyfert-like, redder when brighter behavior.

\section{Broad-band Properties}
In addition to the combination of the X-ray spectral and temporal results,
important insights into the AGN central engine can be obtained by analyzing
the broadband properties, such as SEDs, broadband spectral indices, and 
radio loudness parameters. 

Indeed, the modeling of SEDs 
ranging from the radio to the $\gamma$-ray energies has proved to be one of the
most effective tools to investigate the physics of blazars (e.g., Ghisellini
et al. 2002). The study of SEDs has also been vastly applied to investigate the
nature of the accretion flow in AGN and specifically to discriminate between
radiatively efficient and inefficient scenarios (e.g., Quataert et al. 1999).
Finally, the study of broadband SEDs has been frequently applied to the
scaled-down version of the AGN, the Galactic black hole (GBHs), to investigate
the jet contribution at X-rays in the low/hard spectral state (e.g., Markoff 
et al. 2003). However, despite the much higher signal-to-noise of the GBH data,
the X-ray jet dominance is  still matter of strong debate (Zdziarski et al. 
2003). 
The fact that the discussion is still open for GBHs, the objects 
with the highest S/N and broadest energy coverage, along with
the spectral degeneracy already discussed in $\S$4,
indicates that not even SED studies can be considered exhaustive diagnostics
when used in isolation.

In general, two additional problems plague the 
investigation of AGN SEDs: 1) the relative contributions from different 
energy bands (and hence the overall SED) are known to evolve in time,
casting doubts on the results obtained from non-simultaneous SED studies; 
and 2) the
poor (or lack of) coverage of the UV band, which is thought to map the 
emission from the accretion disk and thus of crucial importance for
AGN studies (Malkan 1983).
 
\xmm, with the EPIC cameras and the OM operating at the same time, offers
the ideal solution to these problems since it covers simultaneously the
0.2--10 keV X-ray range and the optical/UV band.

\subsection{SEDs}
The intrinsic shape of the optical-UV continuum in AGN is still a
matter of debate, because there is no general consensus on the extinction 
corrections that need to be applied.
For completeness, we have tried different extinction curves, including
the Galactic extinction proposed by  Cardelli et al. (1989), the
reddening proposed by Gaskell et al. (2004), and the Small Magellanic
Cloud (SMC) type extinction (Pr\'evot et al. 1984). The results from 
the different extinction laws are consistent with each other 
within 10--20\%. In the
following, we will make use of the de-reddened values obtained with a
SMC-type extinction, since it proved to work better than the other
extinction curves on a large sample of
AGN  from the Sloan Digital Sky Survey (Hopkins et al. 2004).
We also applied extinction corrections by assuming the
absorption column density derived from the X-ray spectral fits, $N_{\rm H}$,
and converting it into optical extinction with the relation $A(V)/N_{\rm H}=
5.3\times 10^{-22}$ (Cox \& Allen 2000). 

The observed magnitudes 
and the de-reddened values obtained
using the SMC-type extinction and  $N_{\rm H}$ from the X-ray analysis 
are summarized in
Table 7. For 3C~88 the values should be considered only as suggestive, since
only the U filter yielded a value for the magnitude. The same value was used for
the other 2 filters only to allow a direct comparison with the other targets
in the 3 optical/UV bands. In addition, unlike the other sources, the U image
of 3C~88 shows a diffuse and uniform emission without any clear enhancement
in the central region.
This may indicate either that the nuclear optical/UV emission is completely
absorbed by the high column density suggested by the $N_{\rm H}-\Gamma$ contour
plots obtained from the X-ray spectroscopy, or that
the nuclear emission is intrinsically too faint to be detectable.

Figure~\ref{figure:fig7} shows the SEDs of the 4 WLRGs from the
radio to the X-rays. Optical/UV and X-ray data are simultaneous, whereas the 
radio data are taken from the literature. The two sets of the optical/UV data 
for \ngc, \pks, and \3c4 correspond to the SMC-type and X-ray extinction 
corrections, respectively.
\begin{table*} 
\caption{Broadband Properties}
\begin{center}
\begin{tabular}{lccccccccccc}
\hline
\hline
\noalign{\smallskip}
Source &&   \multicolumn{2}{c}{$\alpha_{OX}$} &&  \multicolumn{2}{c}{$\log R_{UV}$}  &  $\log R_X$  \\
\noalign{\smallskip}
  && SMC & X-ray &&  SMC & X-ray &   \\
\noalign{\smallskip}
\hline
\noalign{\smallskip}
\noalign{\smallskip}
  NGC~1692   && -0.92 & -1.39 && 4.33 & 3.07 & -1.27\\
\noalign{\smallskip}
\hline
\noalign{\smallskip}
 PKS~0625-35  && -1.17 & -1.26 && 2.94 & 2.71 & -1.98\ \\
\noalign{\smallskip}
\hline
\noalign{\smallskip}
3C~88      && -1.38 & $\dots$ && 3.26 &  $\dots$ & -1.50\ \\
\noalign{\smallskip}
\hline
\noalign{\smallskip}
3C~444     && -0.85 & -0.94 && 2.60 & 2.38 & -3.77\   \\
\noalign{\smallskip}
\hline
\hline
\end{tabular}
\end{center}
\footnotesize
{\bf Columns Table 8}: 1= Source name. 2= $\alpha_{\rm OX}$   computed assuming a SMC-type extinction 
curve.
3= $\alpha_{\rm OX}$  computed assuming $A(V)/N_{\rm H}=5.3\times 10^{-22}$ with $N_{\rm H}$ derived from
the X-ray spectral fitting.
4= UV radio loudness $R_{\rm UV}= L_\nu({\rm 5~GHz)}/L_\nu({\rm 2500~\AA})$ computed assuming a SMC-type
extinction curve.
5=  UV radio loudness computed assuming $A(V)/N_{\rm H}=5.3\times 10^{-22}$ with $N_{\rm H}$ derived from
the X-ray spectral fitting.
6= X-ray radio loudness $ R_{\rm X}=\nu L_\nu({\rm 5~GHz)}/L_{\rm 2-10~ keV}$.
\label{tab8}
\end{table*}

\begin{figure}
\begin{center}
\includegraphics[bb=35 30 369 305,clip=,angle=0,width=9cm]{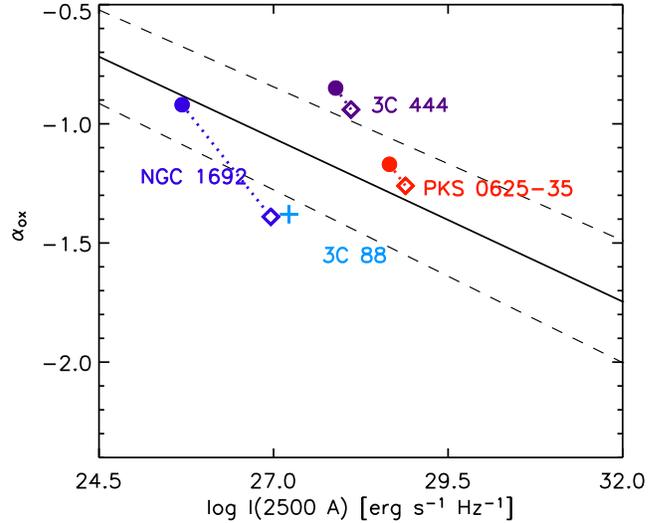}
\end{center}
\caption{$\alpha_{\rm OX}$ plotted versus  2500 \AA\ monochromatic luminosity.
The black continuous line corresponds to the best-fit linear regression found 
by Steffen et al. (2006) and the dashed lines account for the uncertainties in
the fit. Filled circles correspond to values obtained assuming Galactic 
extinction correction, whereas open diamonds correspond to values obtained
assuming the extinction provided by the X-ray spectral fitting. For 
completeness, also 3C~88 has been shown, however its $\alpha_{\rm OX}$ 
value should
be regarded as highly uncertain (see text for details). }
\label{figure:fig8}
\end{figure}

The nuclear emission of weak radio galaxies is commonly ascribed to a
combination of an ADAF (or more generally a RIAF) component and the
non-thermal continuum expected from the base of the relativistic jet,
with the two emissions possibly dominating in different energy bands
(e.g. Quataert et al. 1999). On the other hand, it is also conceivable, 
as discussed in the Introduction, that optical emission from LINERS is 
dominated by a standard accretion flow.

The SEDs of the four sources analyzed in this work share some
similarities (the optical seems to dominate the bolometric output) but
are also different, especially in the X-ray band. A detailed modeling
of the overall emission is beyond the scope of this paper (for a
recent work see e.g. Wu et al. 2007). As an example, we report a
possible modeling of the emission from PKS 0625-35, whose overall
shape is rather suggestive of a dominance of the non-thermal
synchrotron continuum of the jet from the radio to the X-ray
band. Evidence for the presence of a bright jet also comes from VLBA
observations (e.g. Venturi et al. 2000).

The model assumes a spherical emission region (with radius $R$), in
motion with bulk Lorentz factor $\Gamma $ at an angle $\theta $ with
respect to the line of sight (the latter parameters are combined in
the Doppler relativistic factor $\delta$). The source is filled by
tangled magnetic field with intensity $B$ and by relativistic
electrons assumed to follow a smoothed broken power-law energy
distribution with normalization $K$, extending from $\gamma _{min}$ to
$\gamma _{max}$ and with indices $n_1$ and $n_2$ below and above the
break located at $\gamma _b$. This model is clearly rather
simplified. As discussed by Ghisellini et al. (2005) we expect that
subpc scale FRI jets have a complex structure: The emission of
radio-galaxies, in which the jet is misaligned, is possibly dominated
by the lateral walls (the ``layer'') of the jet, surrounding the
faster and more beamed spine responsible for the emission of blazars,
the aligned version of radiogalaxies.

The curve reported in Fig~\ref{figure:fig7} assumes the following values: 
$R=6\times
10^{16}$ cm, $\delta =4$, $B=0.3$ G, $K=3\times 10^2$ cm$^{-3}$,
$\gamma _{min}=1$, $\gamma _b=3\times 10^4$ , $\gamma _{max}=2\times
10^6$, $n_1=2$, $n_2=4$. The power carried by the jet, calculated
assuming the presence of 1 proton per emitting electron (e.g. Maraschi
\& Tavecchio 2003) is $P_j=2.5\times 10^45$ erg/s. Note that the curve
underestimates the radio flux since in this band the emission is
self-absorbed. As generally assumed in the case of sub-pc scale jets,
the emission in the radio band is probably the sum of the emission
from larger portions of the jet.

\subsection{UV -- X-ray constraints}
Recently, Steffen et al. (2006) showed for a sample of optically selected
(Seyfert-like) AGN that the broadband spectral index $\alpha_{\rm OX}$ is 
strongly correlated with the UV monochromatic luminosity, suggesting the
existence of AGN spectral evolution with the luminosity. Maoz (2007)  
performed a similar study for a sample of unobscured LINERs showing that they 
follow a similar trend. Here, we apply the same approach to our sample 
by computing the spectral index, $\alpha_{\rm OX}=
\log(l_{\rm 2500\AA}/l_{\rm 2keV})/\log(\nu_{\rm 2500\AA}/\nu_{\rm 2keV})$
(Tananbaum et al. 1979),
and plotting it versus the UV monochromatic luminosity. The values of the
flux at 2500\AA\ have been obtained by converting the flux measured in the
UVM2 band (2310\AA) assuming a typical slope of 0.7 ($f_\nu\propto\nu^{-0.7})$.

The values of $\alpha_{\rm OX}$, 
computed assuming the two extinction prescriptions discussed above, are 
summarized in Table 8 and plotted in Figure~\ref{figure:fig8}, superimposed to
the best-fit linear regression found by Steffen et al. (2006).  
\3c4, \pks, and \ngc\ show a broad agreement with the best-fit linear 
correlation suggesting a similarity (or at least a smooth transition) with
Seyfert-like objects. On the other hand, no firm conclusions can be drawn
for the remaining 3C~88, due to the large uncertainty
of its UV luminosity.

\subsection{Radio constraints}
\begin{figure}
\begin{center}
\includegraphics[bb=35 30 369 305,clip=,angle=0,width=9cm]{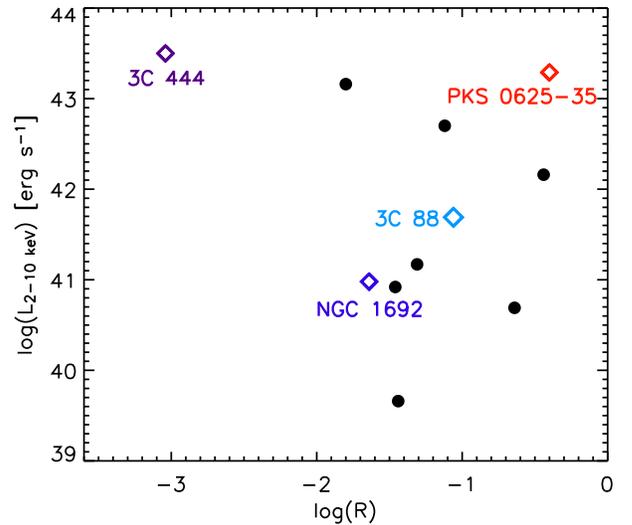}
\end{center}
\caption{2--10 X-ray luminosity associated with the power-law component 
plotted against the radio core dominance. Open diamonds are the 4 sources
analyzed in this work. Filled circles refers to weak-line radio 
galaxies/LINERs analyzed by Rinn et al. 2005.}
\label{figure:fig9}
\end{figure}
 Important constraints on the nature of radio galaxies/LINERs can be
obtained by combining our \xmm\ data with archival radio information,
keeping in mind that the data are not simultaneous. 

For example, a possible way to explore the origin of the X-ray luminosity
(associated with the power-law component) in radio galaxies is to plot it
versus the radio core dominance, $\Re=L_{\rm core}/L_{\rm lobe}$, which is
commonly used as indicator of the jet inclination (the larger $\Re$ the smaller
the inclination angle). According to the classification proposed by Baker
\& Hunstead (1995) based on $\Re$, radio-loud objects can be divided into
3 subsets: face-on objects (likely jet-dominated) with $\Re\ge 1$, 
intermediate objects ($0.1 < \Re < 1$), and edge-on objects ($\Re\le 0.1$).
All our sources have relatively low values, as expected for radio galaxies.
Specifically, \ngc, 3C~88, and \3c4\ fall in the edge-on category, whereas
\pks\ is classified as an intermediate object (see last column of Table 1). 

This approach has been used  by Kharb \& Shastri (2004)
to investigate the origin of the optical nuclear emission in a sample of
radio galaxies observed with \hst. The strong positive correlation prompted
these authors to conclude that the optical nuclear emission is strongly
beamed and likely to be related to the unresolved base of the jet.
On the other hand, the same analysis carried out by Rinn et al. (2005)
for the X-ray luminosity of
a sample of radio galaxies/LINERs did not revealed any positive trend
between $L_{\rm 2-10~keV}$ and $R$, suggesting that the X-ray emission
is not beamed and hence more likely to be related to the accretion flow.

Given the common classification of the sources belonging to the sample 
analyzed by Rinn et al. (2005) and ours (all are radio galaxies whose 
nuclei are optically classified as LINERs), we have combined the 2 samples
considering only the most reliable X-ray and radio data. The results,
plotted in Figure~\ref{figure:fig9}, clearly indicate the lack of any
correlation. This is confirmed by a Spearman rank correlation analysis that
yields $\tau=-0.15$ with a chance probability of $\sim$68\%.
  
Further useful information can be provided by the radio loudness parameter.
Originally, the radio loudness, $R_O$, was defined by the ratio of the 
5 GHz to the optical B band flux densities and used to discriminate between 
radio-quiet and radio-loud ($R_O>10$; Kellermann et al. 1989). More recently,
other definitions of radio loudness have been introduced, making use of energy 
bands less affected by absorption ($R_X=\nu L_\nu({\rm 5~GHz)}/
L_{\rm 2-10~ keV}$; Terashima \& Wilson 2003) or with lower
galaxy contribution ($R_{UV}=L_\nu({\rm 5~GHz)}/L_\nu({\rm 2500~\AA})$; 
Maoz 2007) than the B band.

In his analysis of the SEDs of 7 low-power AGN/LINERs, Ho (1999) pointed 
out the
relatively high radio loudness of these objects, suggesting the existence of an
inverse relationship between radio loudness and Eddington ratio. This 
relationship has been confirmed by several observations, which have also
revealed the existence of two parallel trends in the  $R_O$--Eddington ratio 
plot: An upper ``radio-loud'' branch populated by FRI, FRII, and radio-loud
QSO, and a lower ``radio-quiet'' branch with LINERs, Seyfert galaxies and 
radio-quiet quasars (see Sikora et al. 2007 and references therein).

Recently, Panessa et al. (2007) investigated the radio properties of a 
large sample of Seyfert galaxies and LINERs using both $R_O$ and $R_X$.
They confirmed the existence of a strong anti-correlation between radio loudness
and Eddington ratio and empirically determined the threshold between 
radio-quiet and radio-loud objects in the  $\log R_X-\log L_{\rm 2-10 ~keV}$ 
plot to be around $\log R_X\sim-2.75$. Similarly, Maoz (2007), extending the
Sikora et al. (2007) plane to the UV band, showed that the unobscured LINERs 
observed with \hst\ preferentially lie on the radio-quiet branch of the 
$\log R_{UV}-\log L_{\rm UV}$ plot.

We computed $R_X$ and $R_{UV}$ for our sample (the results
are summarized in Table 8). Our results confirm the anti-correlation
between radio loudness and Eddington ratio. As for the location of our
targets in the $\log R_{UV}-\log L_{\rm UV}$ plot, we
found that \3c4\ lies well within the 
radio-quiet region/branch, \pks\ is in between the 2 branches, whereas
\ngc\ and 3C~88 are more consistent with the bottom part of the 
radio-loud branch. However, once more the
UV values for the latter 2 objects are severely affected by the large 
uncertainty of the extinction.

\section{Summary and Conclusions}
We have investigated the nature of 4 WLRGs,
\ngc, \pks, 3C~88, \3c4, using
observations from \xmm, and combining information from the time-averaged
spectral analysis, from the temporal and spectral variability, and from
broadband studies.
The main results can be summarized as follows:

\begin{itemize}

\item The 4 AGN have Eddington ratios $L_{\rm bol}/L_{\rm Edd}$ that span 
2 orders of
magnitude ranging between $\sim 1\times 10^{-5}$ (\ngc) and $2-3\times 10^{-3}$
(\pks\ and \3c4), with 3C~88 having an intermediate value of 
$\sim 2\times 10^{-4}$.

\item The 4 targets are adequately fitted by the same continuum model that
comprises at least one thermal component ($kT\sim 0.65-1.45$ keV)
and a partially absorbed power law. However, the relative contributions of the 
individual spectral components and the photon indices change significantly 
from source to source. For example, \ngc\ and \pks\ have fairly steep 
power-law components ($\Gamma\sim 2.5-2.9$), whereas 3C~88 has a
flat  photon index ($\Gamma\sim 1.1$), and \3c4\ is the only one showing
a typical Seyfert-like $\Gamma \sim1.9$ accompanied by an apparent line-like 
excess around 6.7 keV. If this excess is fitted with a Gaussian model, the 
resulting line is centered around 6.7 keV, 
fairly strong ($EW\sim550$ eV) and possibly broad ($\sigma\sim0.35$ keV).
However, observations
with a better coverage of the hard X-ray band and with longer duration
are necessary to firmly establish the presence of a broad line and 
constrain the line parameters

\item There is evidence only for moderate absorption 
($N_{\rm H}\sim10^{21}$\nh) in \ngc, \pks, and \3c4. On the other hand, the
flat photon index derived for 3C~88 and the $N_{\rm H}-\Gamma$ contour plots
suggest that this source might have a larger column density of the order of
$N_{\rm H}\sim10^{22}$\nh.

\item The results of the variability analysis indicate that \pks\ is 
significantly variable in the 0.2--10 keV and 0.2--1 keV energy bands,
whereas 3C~88 is variable in the soft band only, and \ngc\ is marginally
variable in the soft band. In general, there is a 
suggestive evidence that 
the sources are more variable in the soft band, although the limited 
statistics due to the short observations
does not allow to draw firm conclusions.

\item For \pks, the split of the exposure into two observations
taken 4 months a part, makes it possible to investigate the
spectral variability using the hardness ratio. The spectrum
becomes softer as the source brightens, following the typical 
Seyfert-like spectral behavior.

\item The broadband SED of \pks, comprising 
(non-simultaneous) radio data combined with (simultaneous) UV and X-ray 
measurements, can be adequately fitted with a single synchrotron component 
with little contribution from the putative accretion disk. However, the paucity
of data-points does not allow one to rule out alternative scenarios.

\item  For \3c4, \pks, and \ngc,  the values of $\alpha_{\rm OX}$
plotted versus the UV monochromatic luminosity follow 
the best-fit linear regression found by Steffen et al. (2006)
for Seyfert galaxies. This suggests a similarity (or at least a smooth 
transition) with Seyfert-like objects.
No firm conclusions can be drawn
for the remaining  object 3C~88, due to the large uncertainty
of its UV luminosity. 

\item The lack of a positive correlation between  $L_{\rm 2-10~keV}$
and the viewing angle, parameterized by the radio core dominance, 
suggests that the X-ray emission is not beamed and hence more 
likely to be related to the accretion flow.

\item The study of the X-ray and UV radio loudness indicates that
 \3c4\ lies well within the 
radio-quiet region/branch, \pks\ is in between the 2 branches, whereas
\ngc\ and 3C~88 are more consistent with the radio-loud branch. 

\end{itemize}

Before concluding, we can try to combine the different findings from our
analysis in an attempt of answering the fundamental question on the origin of
the X-ray emission in WLRGs.

Several pieces of evidence seem to favor an accretion-related origin. First,
relatively high values of $L_{\rm X}/L_{\rm Edd}$, as those found for 3 of our
targets, are commonly interpreted as Comptonized emission from
a RIAF, based on studies of LLAGN and GBHs in the low/hard state 
(Yuan \& Cui 2005; Wu et al. 2007). Second, the lack of a positive correlation
in the  $L_{\rm X}-R$ plane (see  Figure 9), suggests that the X-ray emission
associated with the power law component is not beamed and hence unlikely to
be produced by a relativistic jet. Third, the results from the temporal and
spectral variability analyses (more pronounced variability in the soft band and
the spectral softening associated with the source brightening observed in 
\pks) are more consistent with the trends observed in Seyfert galaxies
and at odds with the typical blazar-like behavior. Finally, the time-averaged
spectral results of \3c4\ ($\Gamma\sim 1.9$ and possibly broad \feka), coupled
with its radio-quiet nature and the relatively high luminosity and Eddington
ratio, suggest that not only this source is unlikely jet-dominated but also 
that the accretion mode might be similar to the Seyfert-like one (i.e., 
radiatively efficient).

On the other hand, a significant (or even dominant) contribution from the jet
in the X-ray regime cannot be a priori excluded at least for some objects.
For example, the main properties of \ngc\ --$L_{\rm X}/L_{\rm Edd}\sim8\times
10^{-7}\,,\Gamma\sim2.9$ and strong radio-loudness-- seem to naturally fit
the jet-dominated scenario. Similarly, the SED and the steep $\Gamma$ of \pks,
coupled with the relatively small viewing angle, might be adequately 
interpreted in the framework of jet-dominated models.

In summary, based on the analysis of time-averaged spectra combined with
model-independent information from X-ray temporal and spectral variability, 
and with inter-band information, we cannot derive a general conclusion
valid for the entire sample. In fact, the main conclusion of this work is that 
LINERs represent
a very heterogeneous class. Not only they comprise non-AGN and AGN objects,
not only the AGN class can be divided into radio-quiet and radio-loud LINERs, 
but also the latter subclass seems to encompass quite different
objects. These may  range from intrinsically low-luminosity objects (in terms of
Eddington ratio), to objects potentially strongly absorbed, to brighter 
objects with properties more in line with Seyfert galaxies.
Our results are based on a very small sample,
which needs to be extended to derive more general conclusions.
Nevertheless, it demonstrates
the need of high-quality X-ray observations to disentangle the relative
contribution of the different components in order to
shed some light on the nature of low-power radio galaxies.

\begin{acknowledgements} 
We thank the referee for the
comments and suggestions that improved the clarity of the paper.
We are grateful to Ari Laor, Ramesh Narayan, and Matthew Malkan for 
interesting discussions and useful suggestions.
MG acknowledges support by the XMM-Newton Guest Investigator Program
under NASA grants 200866 and 201101. Funds from the NASA LTSA grant
NAG5-10708 are also gratefully acknowledged.
\end{acknowledgements}


\end{document}